\documentclass[twocolumn]{aastex63}

\usepackage{afterpage}

\usepackage{booktabs} 
 
\usepackage{siunitx} 
\usepackage{amsmath,amstext,color}
\usepackage{gensymb}
\usepackage{natbib}
\usepackage{hyperref}
\usepackage{longtable}
\usepackage{threeparttablex}
\usepackage{booktabs}

\usepackage{color}
\usepackage{graphicx}
\usepackage{amsmath}
\usepackage{amssymb}

\usepackage[]{appendix}

\begin{document}
\title{Probing Kilonova Ejecta Properties Using a Catalog of Short Gamma-Ray Burst Observations}
\correspondingauthor{Jillian Rastinejad}
\email{jillianrastinejad2024@u.northwestern.edu}

\shorttitle{Short GRB Kilonova Limits}
\shortauthors{Rastinejad et al.}

\graphicspath{{./}{figures/}}

\newcommand{\NU}{\affiliation{Center for Interdisciplinary Exploration and Research in Astrophysics and Department of Physics and Astronomy, \\ Northwestern University, 2145 Sheridan Road, Evanston, IL 60208-3112, USA}}

\newcommand{\CCA}{\affiliation{Center for Computational Astrophysics, Flatiron Institute, 162 W. 5th Avenue, New York, NY 10011, USA}}
\newcommand{\Columbia}{\affiliation{Department of Physics and Columbia Astrophysics Laboratory, Columbia University, New York, NY 10027, USA}}

\newcommand{\CfA}{\affiliation{Center for Astrophysics\:$|$\:Harvard \& Smithsonian, 60 Garden St. Cambridge, MA 02138, USA}}

\newcommand{\Bath}{\affiliation{Department of Physics, University of Bath, Claverton Down, Bath, BA2 7AY, UK}}

\newcommand{\GWU}{\affiliation{Department of Physics, The George Washington University, Washington, DC 20052, USA}}

\newcommand{\Steward}{\affiliation{Steward Observatory, University of Arizona, 933 North Cherry Avenue, Tucson, AZ 85721-0065, USA}}

\newcommand{\Leicester}{\affiliation{School of Physics and Astronomy, University of Leicester, University Road, Leicester, LE1 7RH, UK}}

\newcommand{\Radboud}{\affiliation{Department of Astrophysics/IMAPP, Radboud University, 6525 AJ Nijmegen, The Netherlands}}
\newcommand{\Warwick}{\affiliation{Department of Physics, University of Warwick, Coventry, CV4 7AL, UK}}
\author[0000-0002-9267-6213]{J. C.~Rastinejad}
\NU

\author[0000-0002-7374-935X]{W.~Fong}
\NU

\author[0000-0002-5740-7747]{C. D. Kilpatrick} 
\NU

\author[0000-0001-8340-3486]{K.~Paterson}
\NU

\author[0000-0003-3274-6336]{N. R. Tanvir}
\Leicester

\author[0000-0001-7821-9369]{A. J. Levan}
\Radboud\Warwick

\author[0000-0002-4670-7509]{B.~D.~Metzger}
\CCA\Columbia

\author[0000-0002-9392-9681]{E. Berger} 
\CfA

\author[0000-0002-7706-5668]{R.~Chornock}
\NU

\author[0000-0002-9118-9448]{B. E. Cobb}
\GWU

\author[0000-0003-1792-2338]{T. Laskar}
\Bath

\author{P. Milne}
\Steward

\author[0000-0002-2028-9329]{A. E.~Nugent}
\NU

\author[0000-0001-5510-2424]{N. Smith}
\Steward

\begin{abstract}
The discovery of GW170817 and GRB\,170817A in tandem with AT\,2017gfo cemented the connection between neutron star mergers, short gamma-ray bursts (GRBs), and kilonovae. To investigate short GRB observations in the context of diverse kilonova behavior, we present a comprehensive optical and near-infrared catalog of 85 bursts discovered over 2005--2020 on timescales of $\lesssim12$~days. The sample includes previously unpublished observations of 23 bursts, and encompasses both detections and deep upper limits. We identify 11.8\% and 15.3\% of short GRBs in our catalog with upper limits that probe luminosities lower than those of AT~2017gfo and a fiducial NSBH kilonovae model (for pole-on orientations), respectively. We quantify the ejecta masses allowed by the deepest limits in our catalog, constraining blue and `extremely blue' kilonova components of 14.1\% of bursts to $M_{\rm ej}\lesssim0.01-0.1 M_{\odot}$. The sample of short GRBs is not particularly constraining for red kilonova components. Motivated by the large catalog as well as model predictions of diverse kilonova behavior, we investigate modified search strategies for future follow-up to short GRBs. We find that ground-based optical and NIR observations on timescales of $\gtrsim 2$~days can play a significant role in constraining more diverse outcomes. We expect future short GRB follow up efforts, such as from the {\it James Webb Space Telescope}, to expand the reach of kilonova detectability to redshifts of $z\approx 1$.
\end{abstract}



\section{Introduction}

The detection of the binary neutron star (BNS) merger GW170817 began a new era of multi-messenger astronomy by providing the first direct link between gravitational-wave (GW)-detected compact object mergers and multi-band electromagnetic transients \citep{Abbott+17a,Abbott+17c}. The nearly simultaneous gamma-ray burst (GRB), GRB\,170817A \citep{Abbott+17b,Goldstein+2017,Savchenko+17}, linked BNS mergers to short GRBs, cosmological transients with durations $\lesssim2$ seconds and harder gamma-ray spectra \citep{Norris+84,Kouveliotou+93}. Short GRBs are accompanied by a broadband afterglow originating in a jet launched from the merger remnant \citep{Sari+98,Berger+14}. 
At early times, the optical and NIR light curves of short GRBs are dominated by this rapidly-fading, non-thermal afterglow. 
The ensuing optical and near-infrared (NIR) counterpart to GRB\,170817 and GW170817, AT~2017gfo, also connected BNS mergers with kilonovae \citep{Arcavi+17,Chornock+17,Coulter+17,Cowperthwaite+17,Drout+17,Diaz+17,Fong+17,Gall+2017,Hu+17,Kasliwal+17,Lipunov+17,McCully+17,Nicholl+17,Pian+17,Pozanenko+17,Shappee+17,Smartt+17,Soares-Santos+17,Tanvir+17,Utsumi+17,Valenti+17,villar+17}, thermal transients powered by the radioactive decay of $r$-process elements synthesized in neutron-rich ejecta  \citep{lipaczynski98,metzger+10,barneskasen13,Rosswog+14}. As the only well-localized GW merger in the local ($z<0.05$) Universe, this event allowed for extremely well-sampled, multi-band light curves of the kilonova counterpart, enabling us to directly compare AT~2017gfo to observations of its cosmological analogues. 

In addition to BNS mergers, some neutron star-black hole (NSBH) mergers are predicted to produce kilonovae. Thus far, there are only a few potential detections of an NSBH merger, including GW190426\_152155 \citep{GWTC2} and GW190814 \citep{LVC_GW190814}, though none have a confirmed associated kilonova to date (e.g., \citealt{Hosseinzadeh+19,Gomez+19,Morgan+20,Vieira+20,Anand+20,Andreoni+20,Ackley+20,Paterson+20_SAGUARO,Gompertz+20b}). The lack of electromagnetic (EM) emission from GW190814 is consistent with the finding that the primary BH is 23~$M_{\odot}$, as a high mass black hole would not disrupt the NS until it is inside the BH's innermost stable circular orbit, thus accreting all of the ejecta \citep{Foucart2013,FernandezMetzger16}. Kilonovae from BNS and NSBH systems are of particular interest in GW follow-up as they radiate emission relatively isotropically \citep{lipaczynski98,metzger+10}, unlike other signatures of these mergers (e.g., jetted afterglows) which are generally detected at pole-on orientations. 

Theoretical studies have long predicted a diversity of observed kilonova behavior \citep{lipaczynski98,metzgerfernandez14,Kawaguchi+20b}. Model light curves vary in luminosity and color evolution depending on, among other properties, the merger's progenitors (e.g., BNS or NSBH), component masses, and remnant. Final outcomes to a BNS merger span a prompt BH collapse, a short-lived ($\sim100$~ms) differentially rotating hypermassive neutron star (HMNS), a longer-lived ($\gtrsim10$~s) rigidly rotating supramassive NS (SMNS) \citep{Shibata+19} and a stable infinite-lifetime NS \citep{kasen+15,MargalitMetzger19}. If a NS remnant is born with a large magnetic field (a ``magnetar''), it may impart additional energy on the ejected material \citep{MetzgerPiro14,Gao+17,metzger19,Fong+21}. In addition, parameters such as the viewing angle, progenitor mass ratio, spins and magnetic field strengths are all expected to result in observed kilonova light curve diversity \citep{just+15,Metzger+18}.

Concurrent with new discoveries by GW facilities in the nearby Universe, short GRB detections have provided a large population of BNS or NSBH mergers at cosmological distances ($z \approx 0.1-2.2$; \citealt{Berger+14,Fong+17,Paterson+20}). Short GRBs discovered by the Neil Gehrels {\it Swift} Observatory ({\it Swift}; \citealt{Gehrels+04}) have been followed up with deep optical and NIR observations for the past 15 years, primarily to analyze their collimated, non-thermal afterglows \citep{fong+15}. Despite their comparatively large distances, the wealth of optical and NIR follow-up of short GRBs provide unique and valuable information on potential kilonova emission. Indeed, some previous works have used short GRB detections to identify kilonova candidates from late-time excesses of mostly nearby ($z \lesssim 0.5$) bursts \citep{Tanvir+13,berger+13,Jin+15,yang+15,Gao+17,Gompertz+18,troja+19,lamb+19,Rossi+20,Fong+21}. To detect kilonova candidates in optical imaging, these studies have the added challenge of disentangling emission from thermal kilonovae and non-thermal afterglows. Additional work has explored the colors, ejecta masses ($M_{\rm ej}$), velocities ($V_{\rm ej}$), and lanthanide abundances ($X_{\rm lan}$) of candidates \citep{Ascenzi+19}.

If all candidates claimed in previous works are indeed kilonovae, the population demonstrates significant diversity, especially considering they are uniformly observed from a pole-on orientation given the detections of short GRBs. The advantage of inferring kilonova properties from short GRBs is that it removes the effects of viewing angle, implying that the candidates' observed variation is due to the diversity of kilonovae themselves (or potentially jet/kilonova interaction). A focus on deep upper limits allows us to circumvent the additional uncertainties imposed by modeling afterglows while still exploring kilonova diversity, and bring in a much larger population of viable constraining observations.

In this paper, we incorporate the wealth of published short GRB optical and NIR observations as well as new, unpublished data. We add to previous samples by including observations of short GRBs at all redshifts as some optimistic kilonova models are indeed observable at $z>0.5$. The kilonova with the best-sampled light curve, AT~2017gfo, and the wealth of existing kilonova models provide an excellent baseline against which we can compare constraints from a large sample of short GRBs. In Section~\ref{sec:obs} we describe our comprehensive catalog of 85 short GRB observations. In Section~\ref{sec:analysis}, we investigate the diversity of the data by comparing our sample of observations to the light curves of AT~2017gfo, kilonova models and past kilonova candidates. In Section~\ref{sec:kn_params} we use the deepest upper limits in our sample to make constraints on the ejecta mass and velocity parameter space for each event. Finally, in Section~\ref{sec:discussion} we discuss the implications of our analysis and consider approaches to future short GRB kilonova searches. Unless otherwise stated, all observations are reported in AB mag units and corrected for Galactic extinction, and all times since burst ($\delta t$) are reported in the observer's frame. We use a standard cosmology of $H_{0}$ = 69.6~km~s$^{-1}$~Mpc$^{-1}$, $\Omega_{M}$ = 0.286, $\Omega_{vac}$ = 0.714 throughout this work \citep{Bennett+14}.

\section{Observations \& Data Analysis}
\label{sec:obs}

\subsection{Overall Sample Characteristics}

We present optical and NIR observations of 85 bursts discovered from 2005 to 2020 in an updated catalog of short GRB observations that probe kilonova emission properties. We primarily draw from short GRBs ($T_{90} \lesssim 2$ seconds or classified by the \textit{Swift} Burst Alert Telescope (BAT) catalog; \citealt{Lien+16}) detected by BAT on-board {\it Swift}. Four short GRBs in our sample have extended emission in their $\gamma$-ray light curves \citep{Lien+16}. One burst, GRB\,060614, is classified as a long-duration GRB, with $T_{90}\sim100$~s \citep{Parsons+06,Jespersen+20}, but lacked an accompanying supernova ruling out a massive star origin \citep{Gehrels+06,DellaValle+06,GalYam+06}. We consider nine bursts in our sample to be kilonova candidate events.

We incorporate observations of 39 bursts with known redshifts across the full redshift range ($z\approx$ 0.1--2.2). We note that all of these redshifts originate from the most probable host galaxy of each burst. While these associations are all based on low probability of chance coincidence values of $P_{cc}<0.01-0.15$ \citep{Bloom+02}, it is still possible that not all host assignments are correct. We also include data for 46 bursts with unknown distances, for which we assume $z=0.5$\footnote{With the exception of GRB\,180418A, for which we assume $z=1.0$, following the analysis in \citet{RoucoEscorial+20}}, the approximate median redshift of {\it Swift} short GRBs (\citealt{Berger+14,Fong+17,Paterson+20}; see Section~\ref{sec:ratioimpl} for implications of this assumption). We note one burst, GRB\,150424A, for which multiple redshifts have been claimed ($z=0.3$; \citealt{Castro-Tirado+15} and $z=1$; \citealt{Knust+17,Jin+18}). Given the current uncertainty in redshift, we employ $z=0.5$ in our analysis.

We draw optical and near-infrared observations from three sources: (i) unpublished observations which mainly comprise deep upper limits (Section~\ref{sec:unpub}), (ii) deep upper limits from GCN circulars that have not been otherwise published in the literature (Section~\ref{sec:gcnobs}), and (iii) published observations (or reanalyses of published observations) of past short GRBs, some of which have also been included in previous samples (Section~\ref{sec:litsample}). Notably, this work introduces 55 unpublished observations and over 40 deep upper limits from GCNs. In total, we include observations of 85 bursts, representing most of the \textit{Swift} GRBs with $T_{90} \lesssim 2$ seconds or extended emission detected to date.  Observations span $\delta t = 0.003 - 12$ days post-burst (where $\delta t$ is the time since the burst trigger) and the optical and NIR bands ($ugrizYJHK$). All observations are listed in the Appendix, Table~\ref{tab:sample}.

\subsection{Unpublished Observations}
\label{sec:unpub}

We start with new observations of 23 bursts, the majority of which have confirmed X-ray or optical afterglows that provide localization at the $\lesssim1 - 3$\arcsec\ level. Three bursts lack X-ray afterglow detections but our observations of these BAT positions (typically a few arcminutes in uncertainty; Table \ref{tab:sample}) have all or partial optical coverage.

We obtained these observations with the 8m twin Gemini-North and South telescopes, the 6.5m MMT, the 6.5m Magellan/Baade and Clay telescopes, the 4.2m William Herschel Telescope (WHT) and the 3.8m United Kingdom Infrared Telescope (UKIRT). We apply bias, flat-field, and dark corrections when relevant using standard tasks in the IRAF/{\tt ccdproc} package as well as custom python pipelines\footnote{\url{https://github.com/CIERA-Transients/Imaging_pipelines/}}. To align and co-add individual exposures, we use tools in IRAF and astropy (\texttt{CCDProc} and \texttt{Astroalign}; \citealt{tody93,beroiz+20}). We then register the co-added images to an absolute astrometric catalog using \texttt{astrometry.net} \citep{lang10} or standard IRAF tasks to the SDSS or USNO-B1 (for optical images; \citealt{SDSS_DR12, USNO-B}), and 2MASS (for NIR images; \citealt{2MASS}) catalogs.

For most bursts, we obtained a series of observations with the initial set typically at $\delta t \lesssim 3$~days, followed by one to three additional observations spanning timescales of $\delta t = 1-12$~days. For each burst, we perform relative astrometry between the initial image and each of the later sets using \texttt{Astroalign}. To uncover or place limits on transient emission on these timescales, we perform digital image subtraction using the \texttt{HOTPANTS} software package \citep{becker15}.

If an optical or NIR source is detected within the GRB localization regions \citep{Evans+09,Lien+16}, we perform aperture photometry on the residual images to determine the source brightness using the IRAF/\texttt{phot} package. We determine zeropoints by using point sources in common with the SDSS Data Release 12 \citep{SDSS_DR12}, Pan-STARRS DR1 and DR2 \citep{PS1}, 2MASS \citep{2MASS} or USNO-B1 \citep{USNO-B} catalogs. If there are no significant residuals at or near the localization region, we obtain an upper limit from sources detected at 3$\sigma$ significance in the field. As we obtain the limit from the initial image, our values do not include additional noise from the subtraction. In total, we present 52 unpublished upper limits and 3 unpublished detections of 23 bursts, 10 with known redshifts (Table~\ref{tab:sample}).

\subsection{Observations from GCNs}
\label{sec:gcnobs}

\begin{figure*}
\centering
\includegraphics[width=0.8\textwidth]{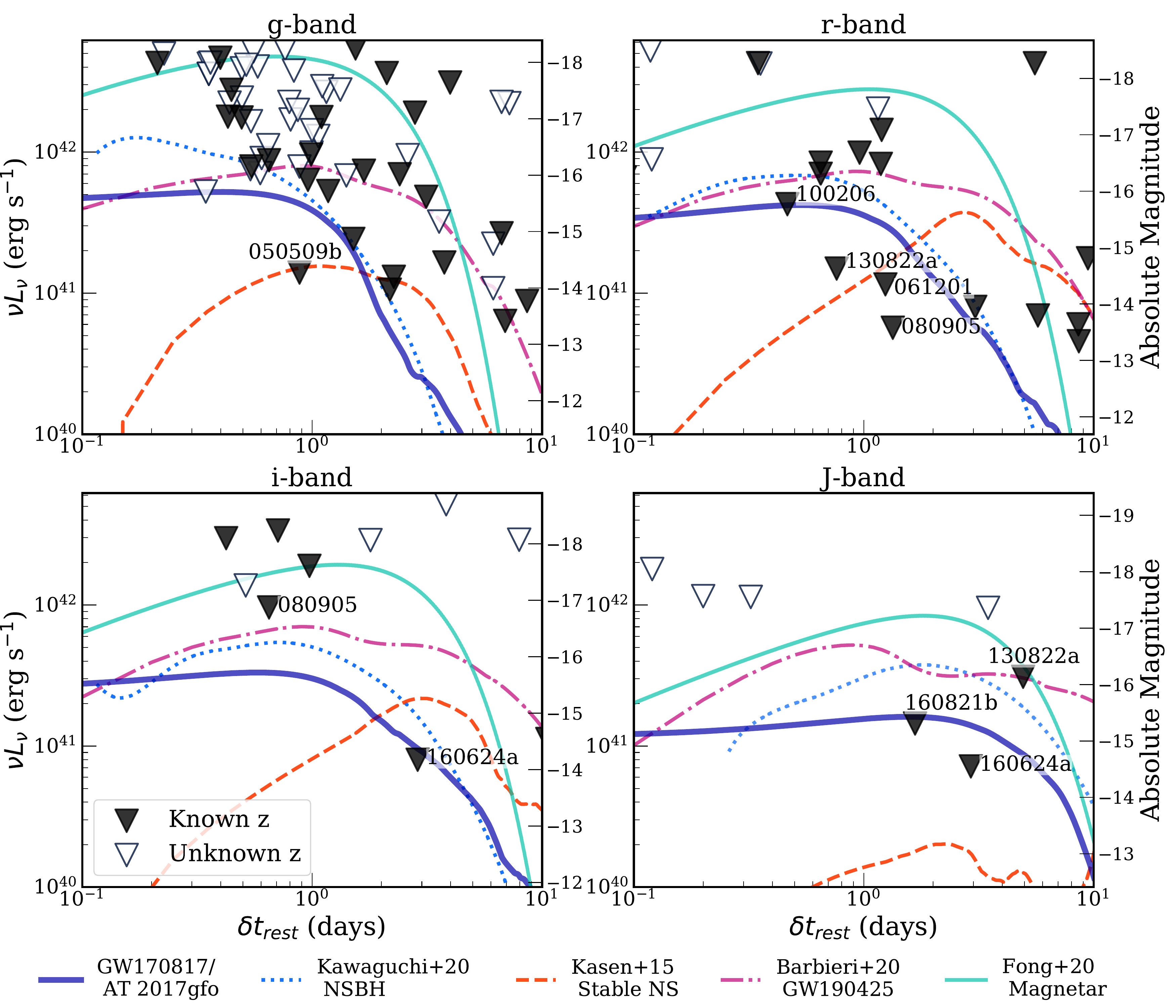}
\vspace{-0.05in}
\caption{Light curves of AT~2017gfo and four diverse kilonova models considered in this work in the $griJ$-bands. Light curves are shown in $\nu L_{\nu}$ (left y-axis) and absolute magnitude (right y-axis) units and rest-frame time in days. In each panel, we show the interpolated, smoothed AT~2017gfo dataset (solid dark blue line, \citealt{villar+17}), an NSBH kilonova model (blue dotted line; \citealt{Kawaguchi+20}), an infinite lifetime NS remnant kilonova model (red dashed line; \citealt{kasen+15}), an extreme optimistic magnetar-boosted kilonova model (turquoise solid line, \citealt{Fong+21}), and a model for an optimistic scenario motivated by GW190425 (magenta dashed-dotted lines; \citealt{Barbieri+20}). Black and open triangles denote upper limits of short GRBs with known and unknown redshifts, respectively, in the respective rest-frame bands at $\delta t_{\rm rest} = 0.1--10$~days. The most constraining limits of bursts with known redshifts are labeled. The largest number of constraining limits appear in the $g$-band (top left panel), typical of afterglow searches.
\label{fig:ratio_lcs}}
\end{figure*}

We search the GCNs of all short GRBs detected between 2013 and 2020 for constraining upper limits to ensure that our catalog covers all \textit{Swift} short GRBs to date (2005--2013 is covered by the literature sample in Section~\ref{sec:litsample}). Considering that the range of peak kilonova luminosities based on models is $\nu L_{\nu} \approx 1 - 5 \times 10^{42}$ erg~s$^{-1}$ ($\sim10$ times the peak luminosity of AT~2017gfo; see Figure~\ref{fig:ratio_lcs}), we only include data that could feasibly fall into the range of kilonova luminosities, or fainter than a given depth, corresponding to $griz \gtrsim 22.5$~mag and $JHK \gtrsim 21.0$~mag (AB system) at $z \approx 0.25$. We choose $z \approx 0.25$ as 80\% of known redshifts in our sample fall above this cut. Observations significantly brighter than this are not included as they are very likely dominated by afterglow emission. In total, we collect deep optical and near-infrared limits of 18 bursts, 7 with confirmed redshifts, from the GCNs (Table~\ref{tab:sample}).

\subsection{Published Observations}
\label{sec:litsample}

To incorporate the full sample of constraining short GRB observations, we undertake a literature search of all relevant short GRBs since 2004. The majority bursts in this sample were discovered by \textit{Swift}-BAT, with the exception of two bursts found by the \textit{INTEGRAL} (GRB\,131224A; \citealt{Mereghetti+03}), \textit{Fermi} and \textit{Suzaku} telescopes (GRB\,140619B; \citealt{fermilat17,Yamaoka+06}). We begin by gathering observations from the catalog presented in \cite{fong+15}. In addition, we retrieve {\it Hubble Space Telescope} ({\it HST}) observations from the archive for GRB\,160624A also published in \citealt{O'Connor+21}). We also collect optical and NIR observations of recently published (2015--present) short GRBs at $\delta t \lesssim 12$ days. 

Finally, we gather relevant detections and upper limits of short GRBs for which kilonovae have been claimed. These kilonova candidate bursts\footnote{We note that \citet{Gao+17} also claim that GRB\,061006 had a detected kilonova. However, following \citet{Rossi+20}, we do not include this candidate in our sample as optical detections at $\delta \gtrsim 1$~day are likely dominated by host galaxy light \citep{D'Avanzo+09}} are GRBs\,050709 \citep{jin+16}, 050724A \citep{Gao+17}, 060614 \citep{yang+15}, 070714B \citep{Gao+17}, 070809 \citep{Jin+20}, 130603B \citep{berger+13,Tanvir+13}, 150101B \citep{troja+18}, 160821B \citep{lamb+19, troja+19} and 200522A \citep{Fong+21}. Though a magnetar-powered kilonova has been claimed for GRB\,080503 \citep{Perley09,Bucciantini+12}, we do not include this burst as a kilonova candidate due to the uncertainty in the burst's redshift and thus optical luminosity \citep{Perley09,FongBerger13}. Assuming $z=0.5$ for GRB\,150424A, we also find luminous ($\gtrsim 10^{41}$ erg s$^{-1}$) optical and NIR detections beyond afterglow-length timescales ($\delta t \gtrsim 5$ days), motivating us to include these observations for completeness (although we do not consider it a kilonova candidate). Including these observations in our literature sample allows us to compare diverse models to the landscape of short GRBs with optical or NIR excesses that have been interpreted as kilonovae. 

Though we apply the analysis in Section~\ref{sec:analysis} to all observations from \citet{fong+15} and the literature, we present only (i) detections that have been interpreted as kilonovae, (ii) deep ($\gtrsim 21.5$~mag) upper limits (see justification of GCN upper limits magnitude cutoff; Section~\ref{sec:gcnobs}) and (iii) low-luminosity afterglow detections (less than the corresponding luminosity of AT~2017gfo; $\nu L_{\nu} \lesssim 5 \times 10^{41}$ erg s$^{-1}$) in Table~\ref{tab:sample}.

\section{Comparisons Between Short GRB Luminosities and Kilonova Models}
\label{sec:analysis}

We now explore our short GRB observational catalog in the context of an AT~2017gfo-like kilonova and three kilonova models. Our sample is fairly heterogeneous, spanning a range of observing bands, redshifts, and timescales. To enable direct comparisons of our observations to both AT~2017gfo and kilonova models, we use the bursts' redshifts to determine luminosities (or luminosity limits), approximate rest-frame bands and rest-frame times after the burst ($\delta t_{\rm rest}$). We apply the following analyses to all observations in Table~\ref{tab:sample}. As all observations were obtained as follow-up to short GRBs, we assume an approximately fixed viewing angle.

\subsection{Comparison to AT~2017gfo}
\label{sec:at2017gfo_comparison}

\begin{figure*}
\centering
\includegraphics[width=0.7\textwidth]{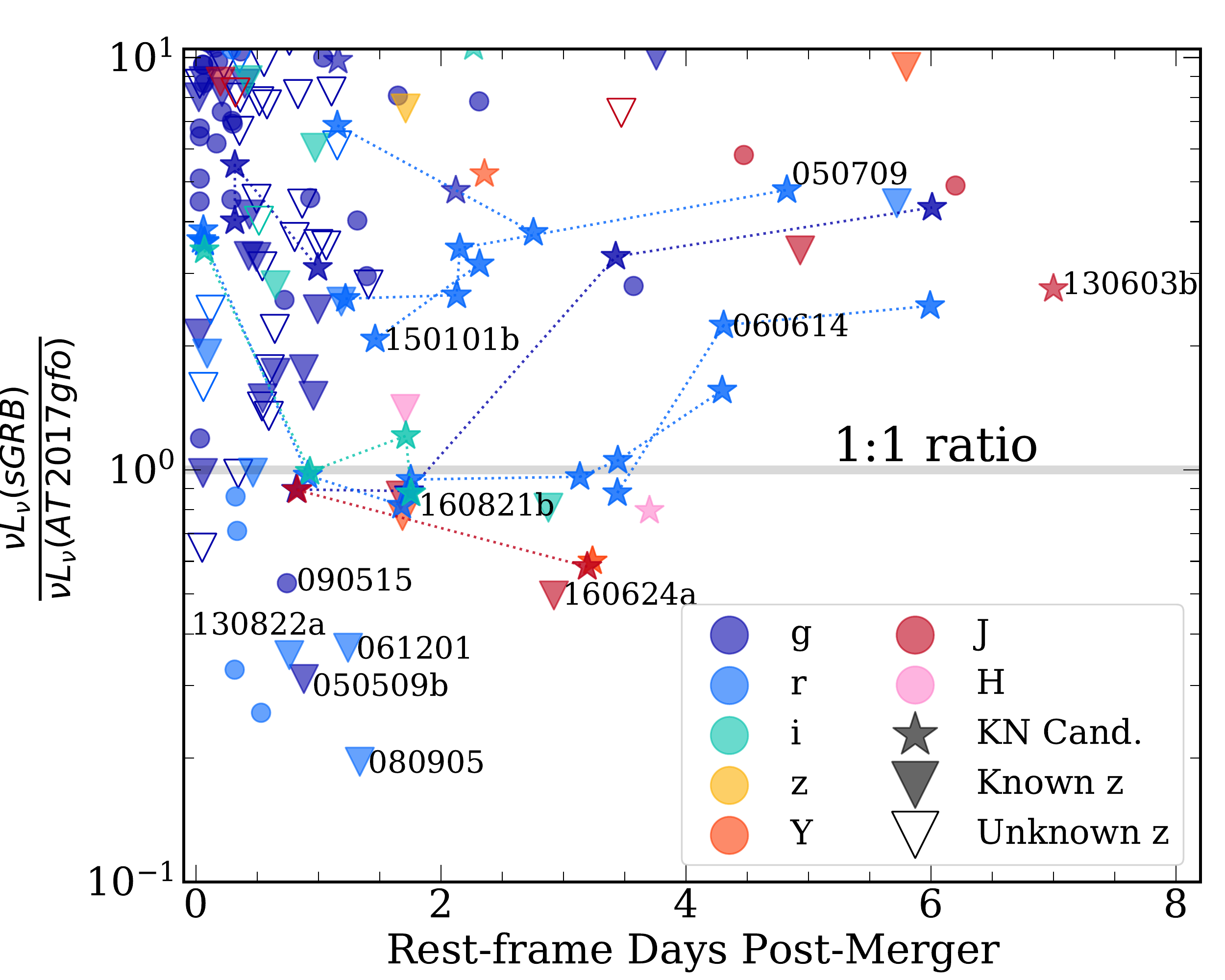}
\caption{The ratios, $\mathcal{R}$, of short GRB observations to the AT~2017gfo light curves, matched in rest-frame time and approximate rest-frame band, in $\nu L_{\nu}$ units (see Section~\ref{sec:at2017gfo_comparison} for ratios method). Filled triangles represent upper limits of bursts with known redshift and unfilled triangles show upper limits of bursts with unknown redshift. Only the deepest upper limit of each short GRB is shown. Circles denote afterglow detections, and stars represent kilonova candidate detections. For kilonova candidates with multiple detections in a single band, stars are connected with dotted lines. The color of each symbol corresponds to the approximate rest-frame band (spanning the $grizyJH$-bands) of each observation. Deep upper limits of 10 bursts, spanning $\delta t_{\rm rest} = 0.05 - 3$~days and the $griyJ$-bands, probe luminosities below AT~2017gfo. We label the GRB names of markers denoting some upper limits with $\mathcal{R} \lesssim 0.8$, relevant kilonova candidates, and low-luminosity afterglows. The gray horizontal line represents a 1:1 ratio against which each short GRB observation can be independently compared. The majority of kilonova candidate detections, including those of GRBs\,050709, 060614, 130603B, 150101B and 200522A have $\mathcal{R} \gtrsim 1$ values.}

\label{fig:8017_ratio}
\end{figure*}

First, we compare our sample to the multi-band light curves of AT~2017gfo at $d_{L} \approx 40.7$~Mpc \citep{Cantiello+18}. The published multi-band dataset of the AT~2017gfo is well-sampled for $\delta t_{\rm rest} \lesssim 12$ days and is compiled in \citealt{villar+17} with observations from \citealt{Andreoni+17,Arcavi+17,Coulter+17,Cowperthwaite+17,Diaz+17,Drout+17,Evans+17,Hu+17,Kasliwal+17,Lipunov+17, Pian+17, Pozanenko+17, Shappee+17, Smartt+17, Tanvir+17,Troja+17, Utsumi+17,Valenti+17}. In each of the $grizyJHK$-bands, we linearly interpolate to 1-hour bins to create finer sampling and smooth with a Savitsky-Golay filter with a time resolution of $\sim1-6$~days, producing master light curves of a AT~2017gfo-like kilonova. We show our interpolated and smoothed AT~2017gfo light curves as blue solid lines in Figure~\ref{fig:ratio_lcs}.

We calculate the ratio, $\mathcal{R}$, of the luminosity of each observation to that of the AT~2017gfo-like light curves in the nearest rest-frame band and time, where $\mathcal{R} \equiv \nu L_{\nu}$(sGRB)/$\nu L_{\nu}$(AT~2017gfo). Due to the large apparent redshift of sGRBs at cosmological distances, $\nu$(sGRB) is approximately but not exactly $\nu$(AT~2017gfo). This method effectively normalizes each short GRB observation by the smoothed light curves, enabling us to directly compare detections and limits across different bands and timescales. We note that this is an approximation, used in the context of a large sample. To quantify the potential uncertainty, we use a model of the AT~2017gfo SED discussed in \citet{kasen+17}. The model SED allows us approximate the luminosity of AT~2017gfo at the true rest-frame wavelength of each short GRB observation at $\delta t < 8$~days and compare this value to that used in our analysis. We find that the median errors of our AT~2017gfo luminosities due to this transformation and across all bands range from 3\%--37\%. The largest median uncertainties are of rest-frame $g$-band observations, likely due to the rapid SED evolution of AT~2017gfo in the bluest bands. We note that the alternative approach to our ratio method, modeling the spectral evolution of AT~2017gfo, would also add uncertainties to our $\mathcal{R}$ values. In Figure~\ref{fig:8017_ratio}, we plot $\mathcal{R}$ as a function of rest-frame time, $\delta t_{\rm rest}$.

This analysis increases the number of short GRBs whose kilonovae must be less luminous than AT~2017gfo. The constraining upper limits span $\delta t_{\rm rest}\approx 0.05-3$~days and are mostly in the bluer rest-frame bands, a product of typical optical afterglow searches and the required shifting of rest-frame-equivalent bands to bluer wavelengths.
From their catalogs of $<40$ short GRBs, previous works found six bursts (GRBs\,050509B, 051210, 061201, 080905A, 100206 and 160624A) which have deep upper limits able to rule out a AT~2017gfo-like kilonova \citep{Gompertz+18,Rossi+20,O'Connor+21}. We corroborate these findings, except for GRB\,051210 for which we consider a higher redshift estimate than previous studies ($z\gtrsim1.55$; \citealt{Berger07}), thus yielding significantly less constraining limits. We find an additional five short GRBs with upper limits lower than or nearly comparable to an AT~2017gfo-like kilonova ($\mathcal{R}\lesssim 1$; GRBs\,130822A, 150120A and 160821B with known redshifts and GRBs\,080503 and 140516 assuming $z=0.5$). However, GRB\,160821B is a known kilonova candidate \citep{lamb+19,troja+19}, highlighting differences in the luminosities and colors as compared to AT~2017gfo.

In the context of the handful of short GRBs with claimed kilonovae in the literature (GRBs\,050709, 050724A, 060614, 070714B, 070809, 130603B, 150101B, 160821B and 200522A; \citealt{Tanvir+13,berger+13,yang+15,jin+16,Gao+17,troja+18, Jin+18,lamb+19,troja+19,Jin+20, Fong+21}), Figure~\ref{fig:8017_ratio} demonstrates that their late-time excesses ($0.5<\mathcal{R}<10$)  exhibit significant diversity in their luminosities and colors. We reconfirm the findings of previous works that the detections of most kilonova candidate bursts have values $\mathcal{R} \gtrsim 1$ in the rest-frame optical band. \citep{Gompertz+20,Rossi+20,Fong+21}. More specifically, only GRBs\,050709, 060614, 130603B and 160821B had some detections at $\delta t_{\rm rest} > 2$ days with comparable luminosities to AT~2017gfo ($\mathcal{R} \approx 0.8-3$). The remaining kilonova candidates had more luminous detection ratios $\mathcal{R} \gtrsim 3$ at $\delta t_{\rm rest} > 2$ days (GRBs\,050724A, 150101B, 200522A) or detections only at early times $\delta t_{\rm rest} < 1.5$ days (GRBs\,070714B, 070809), when extricating afterglow emission is particularly difficult.
 
Finally, we reconfirm the sub-AT~2017gfo ($\mathcal{R}<1$) luminosity of the early detections of GRBs\,061201, 080905A and 090515 found previously \citep{Rossi+20}. The detections of most of these bursts can be explained by their proximal distances, as these three bursts all have low redshifts ($z \lesssim 0.4$; \citealt{Stratta+07,Rowlinson+10,Rowlinson+10b}), Taken together, these low-luminosity afterglow detections, brighter kilonova candidate detections, and constraining upper limits demonstrate that there is a wide diversity in kilonovae observed at a pole-on viewing angle. 

\subsection{Comparisons to Pole-on Kilonova Models}
\label{sec:kn_models}

Motivated by the diversity of kilonova models and the variety of BNS or NSBH merger progenitors discovered by GW detectors (e.g., GW190425, GW190814, \citealt{LVC_GW190425, LVC_GW190814}), we extend our comparative analysis to a set of kilonova models from the literature representing a variety of short GRB progenitors and remnants. Each of the models chosen for our comparisons possesses unique observable signatures that can be tied to the mergers' progenitors and remnants. 

In addition to the BNS merger kilonova, GW170817, we employ a NSBH merger kilonova model to represent a second potential progenitor channel for short GRBs \citep{Kawaguchi+20}. In addition, we consider two models of optimistic, if infrequent, outcomes to a BNS or NSBH merger. These are an infinite lifetime NS remnant to a BNS merger \citep{kasen+15} and a massive BNS merger modeled by the parameters of the unusual event, GW190425 \citep{Barbieri+20,LVC_GW190425}. As the detection of a short GRB automatically provides on-axis (or close to on-axis) orientation, we choose models for pole-on orientations when available.

\begin{figure*}
\centering
\includegraphics[width=0.49\textwidth]{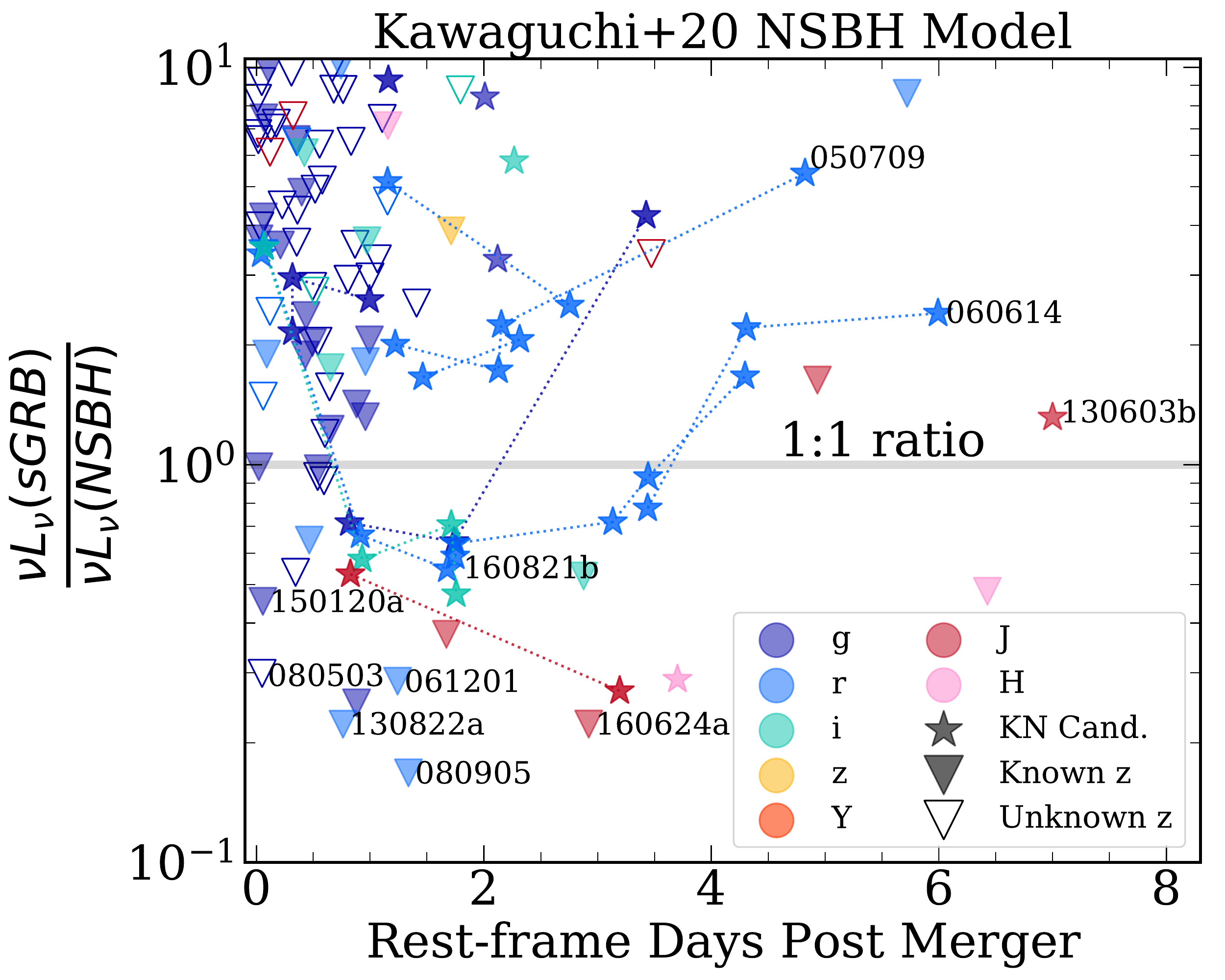}
\includegraphics[width=0.49\textwidth]{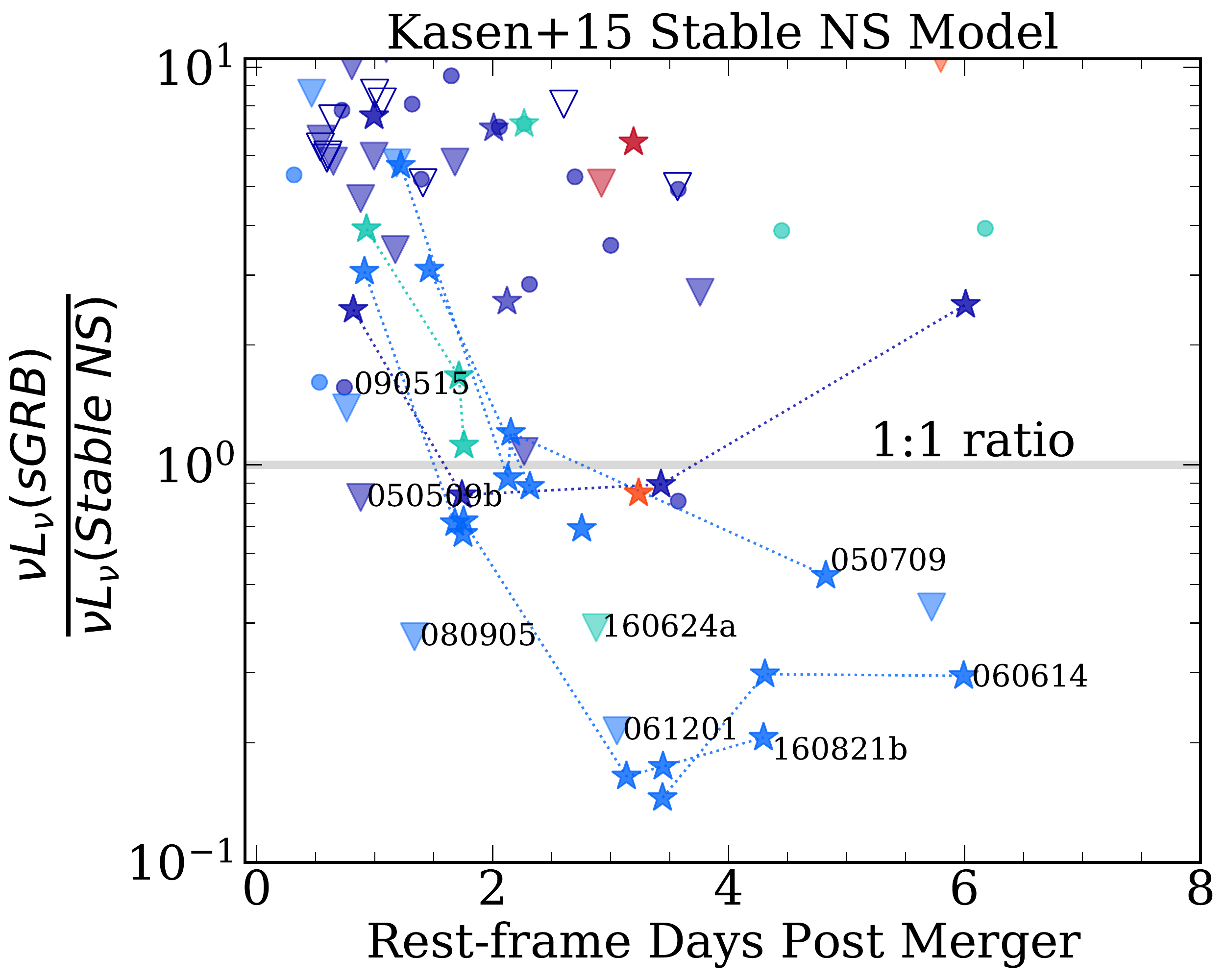}
\includegraphics[width=0.8\textwidth]{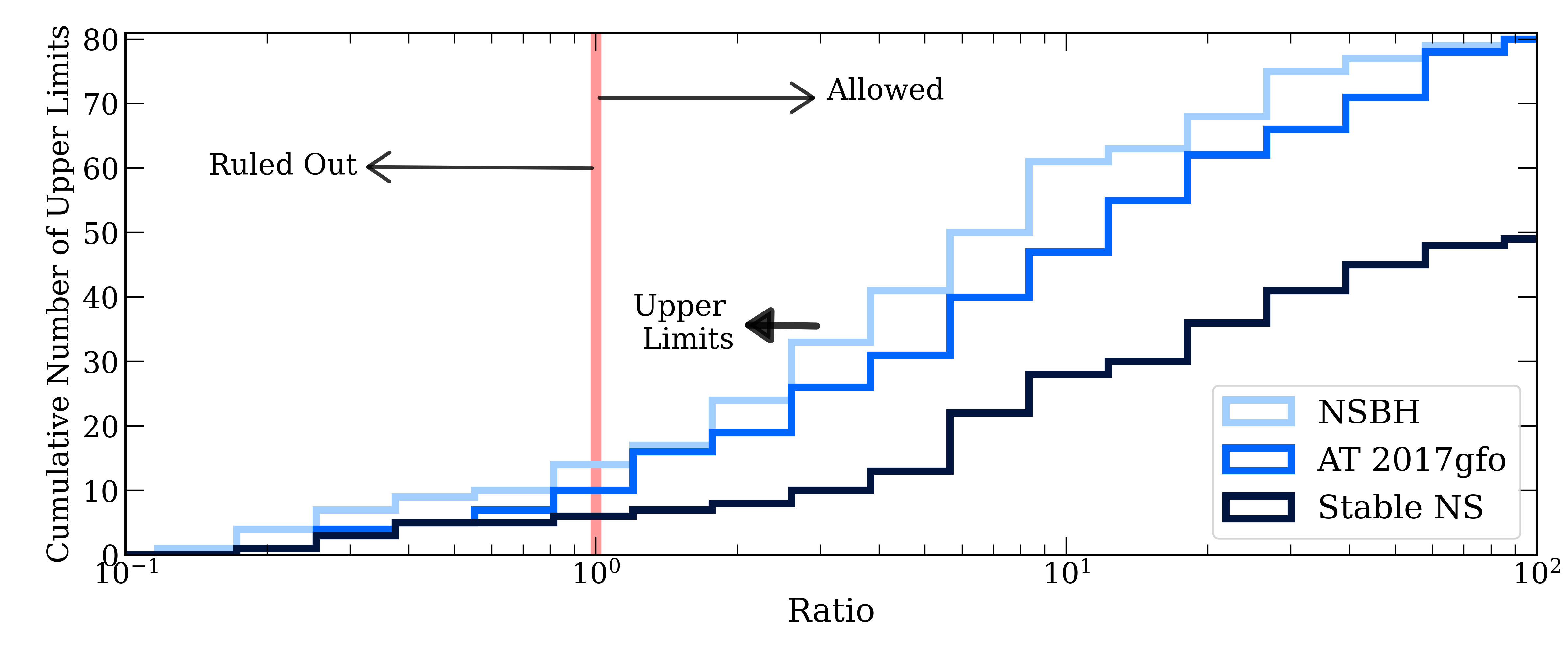}
\vspace{-0.1in}
\caption{{\bf Top Row:} The same as Figure~\ref{fig:8017_ratio} but $\mathcal{R}$ compared to an NSBH model (left; \citealt{Kawaguchi+20}) and an infinite lifetime NS remnant kilonova model (right; \citealt{kasen+15}). Overall our limits have more constraining power on the NSBH model than AT~2017gfo, especially in the NIR bands. Since the infinite NS remnant kilonova model peaks later and at lower luminosities than AT~2017gfo (Figure~\ref{fig:ratio_lcs}), fewer limits rule out this model than either AT~2017gfo or the NSBH case. \textbf{Bottom}: Cumulative distributions of luminosity ratio, $\mathcal{R}$,  values derived from short GRB upper limits of AT~2017gfo (blue), the NSBH model (light blue) and the infinite NS remnant model (dark blue). Only the deepest limit of each burst is included in the distribution. Thus, the cumulative number of upper limits can be thought of as the number of bursts with upper limits that rule out a factor of $\mathcal{R}$ of each model. The vertical line denotes $\mathcal{R}=1$, below which the limit can rule a model out. The NSBH model is ruled out ($\mathcal{R} \lesssim 1$) by 13 upper limits, the largest number of the three models shown.
\label{fig:kn_ratio}}
\end{figure*}

While the overall colors of NSBH merger kilonovae are expected to be redder, the amount of ejecta emitted is variable and depends on several factors including the BH spin, mass ratio and NS properties \citep{Rosswog05,just+15,Kyutoku+15,Foucart+18,Shibata+19}. Indeed, potentially only a small fraction of NSBH mergers produce EM emission \citep{Broekgaarden+21}. Here, a kilonova is comprised of (i) lanthanide-rich dynamical ejecta created by tidal interactions and (ii) post-merger material ejected from the accretion disk formed around the remnant object \citep{Metzger+08,Metzger+09}. For comparison to our short GRB catalog, we choose a representative NSBH model, parameterized by dynamical ejecta $M_{\rm ej} = 0.02 M_{\odot}$ and  $Y_{\rm e} = 0.09 - 0.11$ and post-merger ejecta $M_{\rm ej} = 0.04 M_{\odot}$ and $Y_{\rm e} = 0.1 - 0.3$ \citep{Kawaguchi+20}. This model uses a restricted line list to calculate ejecta opacities (resulting in an error of $\lesssim$ 0.2 mag) and describes a kilonova viewed at pole-on orientation. We note, however, that due to uncertainties in the opacities of highly ionized elements expected in the ejecta at $\delta t < 1$~day, the light curves' luminosity may be artificially enhanced on these timescales \citep{Kawaguchi+20b}. Using supplied $grizJHK$-band models, we calculate $\mathcal{R} \equiv \nu L_{\nu}$(sGRB)/$\nu L_{\nu}$(NSBH Model), and find that upper limits of 13 short GRBs, 9 with known redshifts, at $\delta t_{\rm rest} < 7$ days have $\mathcal{R} < 1$ (Figure ~\ref{fig:kn_ratio}, Panel 1).

We next compare deep upper limits of short GRBs to a model for an infinite lifetime NS remnant \citep{kasen+15}. A long-lived NS remnant will inject energy into the post-merger disk, increasing the amount and the $Y_e$ of disk ejecta material \citep{Dessart+09,perego+14,FahlmanFernandez18,Lippuner+17,Radice+18}. We employ a model at a pole-on orientation ($cos(\theta) = 0.95$) with a disk mass $M_{\rm disk} \sim 0.03 M_{\odot}$ using supplied $grizyJHK$-band light curves \citep{kasen+15}. As these model light curves rise more slowly than AT~2017gfo (Figure~\ref{fig:ratio_lcs}), upper limits at $\delta t \lesssim 2$~days are less constraining. Upper limits of 6 bursts at $0.8 \lesssim \delta t_{\rm rest} < 9$ days are able to rule out this model.

In Figure~\ref{fig:kn_ratio}, we display cumulative distributions of $\mathcal{R}$-value upper limits with respect to the data and two models, including
only the minimum $\mathcal{R}$ value of a given short GRB in each distribution. It is clear that of three distributions, our catalog is most constraining for the NSBH model and least constraining for the infinite lifetime NS remnant model.

Finally, we briefly compare to a model for the unusually massive GW-detected BNS merger, GW190425 \citep{LVC_GW190425}, parameterized by the event's component masses \citep{Barbieri+20}. This model assumes a BNS merger employing the DD2 NS equation of state \citep{Hempel+10,Typel+10}, resulting in a kilonova more luminous than AT~2017gfo. However, we note alternate scenarios have been proposed in which the emission is predicted to be much fainter \citep{Foley+20,Kyutoku+20}. Thus, the \citealt{Barbieri+20} model considered here represents an optimistic outcome for this event, especially at times $\delta t \gtrsim 1$~day (Figure~\ref{fig:ratio_lcs}). Subsequently, limits of 14 bursts rule out this model ($\mathcal{R} < 1$).

\subsection{Implications of Ratio Analysis and Assumptions}
\label{sec:ratioimpl}

We first explore our assumption of a fiducial redshift ($z=0.5$, the median value of \textit{Swift} short GRBs) to the bursts with unknown redshifts of our sample. We note that some bursts with unknown redshifts may have very faint or unidentified host galaxies and thus may be more likely to be at redshifts higher than the observed median, while others may exist at large offsets from their hosts, making precise host identifications based on chance alignment difficult. Higher assumed redshifts undoubtedly affect the constraining power of the sample. For instance, assuming $z=0.5$, upper limits of 2 and 23 bursts have values of $\mathcal{R}_{\rm AT~2017gfo} < 1$ and $\mathcal{R}_{\rm AT~2017gfo} < 10$, respectively. If we instead employ $z=1$ (a conservative estimate) for these bursts' redshift, these numbers drop to 0 and 7, respectively. However, these bursts are included for completeness of the catalog, and may have redshifts identified at a later time. Given this caveat, we find that the most meaningful constraints on short GRB kilonovae come from low-redshift bursts, naturally a result of the smaller luminosity distances rest-frame wavelengths. The eight bursts with upper limits of lower luminosities than AT~2017gfo span the redshift range $0.111 \lesssim z \lesssim 0.483$. 

The majority of kilonova candidates are nearby ($z\lesssim 0.5$) and have larger optical luminosities and bluer colors than AT~2017gfo (likely viewed at $\sim$20 degrees off-axis; \citealt{Mooley+18,Ghirlanda+19,Hotokezaka+19,MarguttiChornock20}), which can partially be explained by  Malmquist bias (implying that we must be probing the upper end of the kilonova luminosity function with short GRBs). Viewing angle may also contribute to the bluer colors as, generally, lanthanide-rich dynamical ejecta is concentrated in the equatorial regions, a result of the tidal tails, while lanthanide-poor emission is distributed isotropically, resulting in bluer emission at the poles \citep{sekiguchi+16,wanajo+14,Piro+18,Metzger+18}. Further observations of on- (short GRB) and off-axis (GW-detected) BNS/NSBH mergers will investigate the extent to which viewing angle plays a role in observed kilonova colors.

Turning to NSBH merger kilonovae, which in general are expected to be redder, the few existing NIR limits at $\gtrsim 1.9$~days offer more constraining power ($\mathcal{R}_{\rm NSBH} < \mathcal{R}_{\rm AT~2017gfo}$) than for AT2017gfo-like kilonovae. Moreover, the single {\it HST} F160W detection of GRB\,130603B more closely traces our NSBH model and is a factor of $\approx 3$ in excess of AT~2017gfo, in line with previous assessments that this burst is consistent with a NSBH merger kilonova \citep{Hotokezaka+13,kawaguchi+16}. While there are optical luminosity differences between the AT2017gfo and the NSBH model considered here, they primarily reside at $\delta t \lesssim 1$~day, when the afterglow would dominate over any kilonova emission. There is potentially a wide diversity in NSBH binary spins and mass ratios \citep[as inferred from, e.g., GW systems;][]{GWTC2} that, if true, imply a wider range in luminosities for NSBH kilonova than those of BNS mergers \citep[in the absence of a long-lived remnant;][]{just+15,Shibata+19}. Indeed, these model comparisons highlight the need for deep, NIR-band follow-up at $\delta t \gtrsim 1$~day to distinguish between canonical BNS and NSBH kilonova models following on-axis short GRBs. 

Our catalog is also well-suited to probe a model that predicts bluer overall colors, as 90\% of the upper limits in our sample are in the rest-frame $gri$-bands. In the long-lived NS remnant scenario, which is expected be rare ($\lesssim3$\% of all BNS mergers are expected to result in a stable NS remnant \citealt{MargalitMetzger19}), the kilonova exhibits a slower rise \citep{kasen+15}. Thus, only rest-frame $r$- and $i$-band observations at $\delta t \gtrsim 2$~days are comparatively meaningful, where our catalog is fairly sparse.

The current short GRB catalog offers significant constraining power on high-luminosity kilonova models, such as the ``magnetar-boosted'' model used to explain the luminous kilonova candidate GRB\,200522A \citep{Fong+21}, of which $\sim$50\% of our limits are constraining. Moreover, motivated by the discovery of the high-mass BNS merger GW190425 (total mass of 3.4 $M_{\odot}$; \citealt{LVC_GW190425}), there are a number of models describing the event's kilonova that range in predicted luminosities by a factor of $\approx 25$. These differences depend on assumptions of tidal deformability, component masses, and the survival of an NS remnant \citep{Barbieri+20,Foley+20,Kyutoku+20}. Adopting the optimistic \citet{Barbieri+20} model as a proof-of-concept, 25\% of upper limits in our sample probe lower luminosities than this model.

\section{Constraints on Kilonova Ejecta Properties}
\label{sec:kn_params}

\subsection{Model Grid Descriptions}

We next determine how the short GRB catalog can place quantitative constraints on kilonova ejecta properties. In particular, we focus on constraints for the ejecta properties ($M_{\rm ej}$, $V_{\rm ej}$) as well as composition (electron fraction $Y_{\rm e}$ or $X_{\rm lan}$). We determine that deep upper limits with values $\mathcal{R}_{\rm AT~2017gfo} \lesssim 3$ are capable of ruling out the extreme upper end of kilonova ejecta masses and velocities ($M_{\rm ej} = 0.5 M_{\odot}$ and $V_{\rm ej} = 0.5c$), which we set as the criteria for subsequent analysis. Ultimately, we find upper limits of 14 short GRBs, including 9 bursts with known redshifts, fit this criteria (see Table~\ref{tab:mej}).

Our goal is to translate each observation into a constraint on the $M_{\rm ej}-V_{\rm ej}$ parameter space at fixed values of $Y_{\rm e}$ or $X_{\rm lan}$. For this, we employ two approaches: direct comparison to a grid of models from \citet{kasen+17} (K17), and forward-modeling based on analytical prescriptions in \citet{metzger19} (M19). The K17 models are based on simulations which include time-varying opacity contributions based on atomic data from both the lighter and lanthanide material synthesized in the kilonova ejecta \citep{Kasen+13}.

Superimposed ``blue'' ($Y_{\rm e}, \gtrsim 0.25$)  and ``red'' ($Y_{\rm e} \lesssim 0.25$) component models are often invoked to describe kilonova light curves \citep{villar+17, kasen+17}. As both components are dominated by thermal emission from radioactive decay \citep[as in][]{Arnett82}, the blue, low opacity kilonova light curves peak earlier in the optical band ($\lesssim 1$ day post-merger) while the red, high opacity light curves peak later in the NIR band (1-2 days post-merger; e.g. AT~2017gfo, \citealt{villar+17,kilpatrick+17,Cowperthwaite+17}). We consider each component separately in this analysis.

We retrieve the publicly-available K17 grids of `red' lanthanide-rich ($X_{\rm lan} = 10^{-2}$) `blue' lanthanide-poor ($X_{\rm lan} = 10^{-4}$) and `extremely blue' ($X_{\rm lan} = 10^{-9}$) components\footnote{\url{https://github.com/dnkasen/Kasen_Kilonova_Models_2017}}. An `extremely blue' component is a predicted product of a long-lived NS remnant \citep{Lippuner+17}. The model grids provide a range of $M_{\rm ej}=0.001-0.1$~$M_{\odot}$ and $V_{\rm ej}=0.03-0.3c$ for fixed $X_{\rm lan}$ values. The fine wavelength sampling ($\lesssim 0.04$ microns) allows us to compare each short GRB observation directly to the relevant rest-frame wavelength. However, the $M_{\rm ej}$ and $V_{\rm ej}$ parameter space is coarsely sampled.

Thus, we generate a more finely-spaced light curve grid following the prescriptions in \citet{metzger19} (M19). Our models utilize the {\tt gwemlightcurves} python-based code \citep{Coughlin+19,Coughlin+20}, with updates for opacity based on the electron fraction ($Y_{\rm e}$) from simulations in \citet{Tanaka+19}. We generate observed light curve models at the redshifts of the 14 bursts used in this analysis (Table~\ref{tab:mej}) for a range of ejecta masses and ejecta velocities. We calculate in-band light curves in the observer frame using {\tt pysynphot}, a {\tt python}-based synthetic photometry package \citep{pysynphot}, accounting for all redshift and time dilation effects when the light curves are generated. The result is log-linear grids of 900 light curve models spanning $M_{\rm ej}=0.001-0.5$~$M_{\odot}$ and $V_{\rm ej}=0.03-0.5c$ for $Y_{\rm e}=0.15$ (a `red' component), $Y_{\rm e}=0.40$ (a `blue' component), and $Y_{\rm e} = 0.45$ (an `extremely blue' component). We note that the M19 $Y_{\rm e}$ values do not directly correspond to the K17 values of $X_{\rm lan}$, so we include both models with their respective values for completeness. We note that the lack of time-varying opacities in this method leads to less constraining results (see Section \ref{sec:moddependencies} for a more thorough analysis of the differences between K17 and M19 models), although this allows for finer sampling of the $M_{\rm ej}-V_{\rm ej}$ grid, and extension to larger values.

\begin{figure*}
\centering
\includegraphics[width=\textwidth]{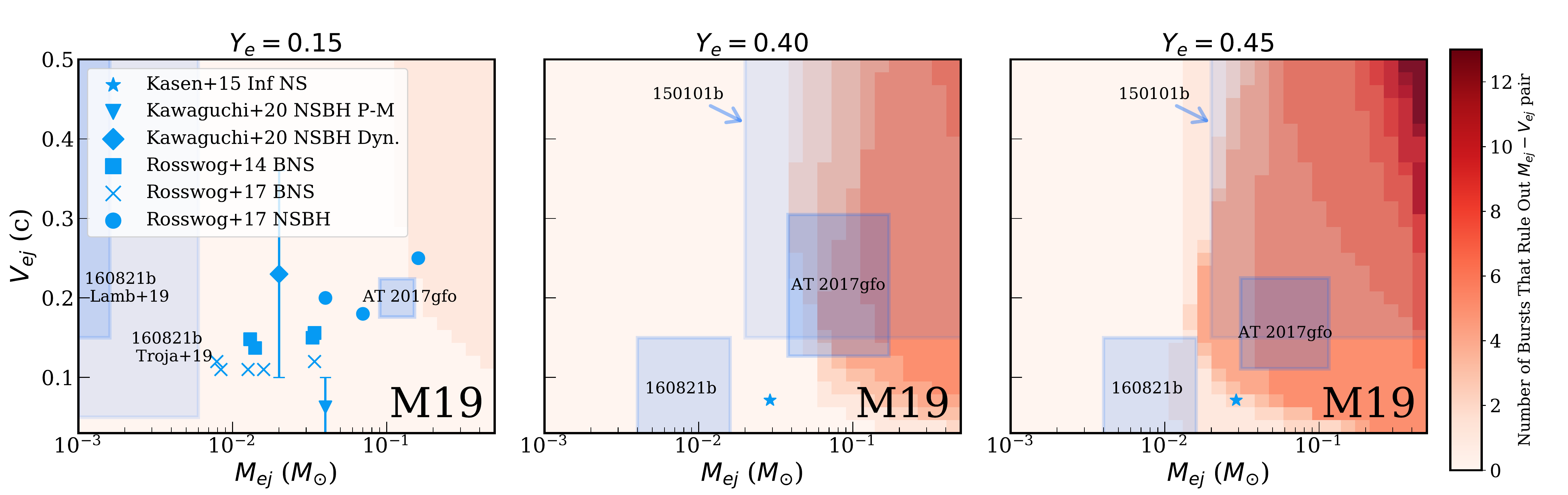}
\includegraphics[width=\textwidth]{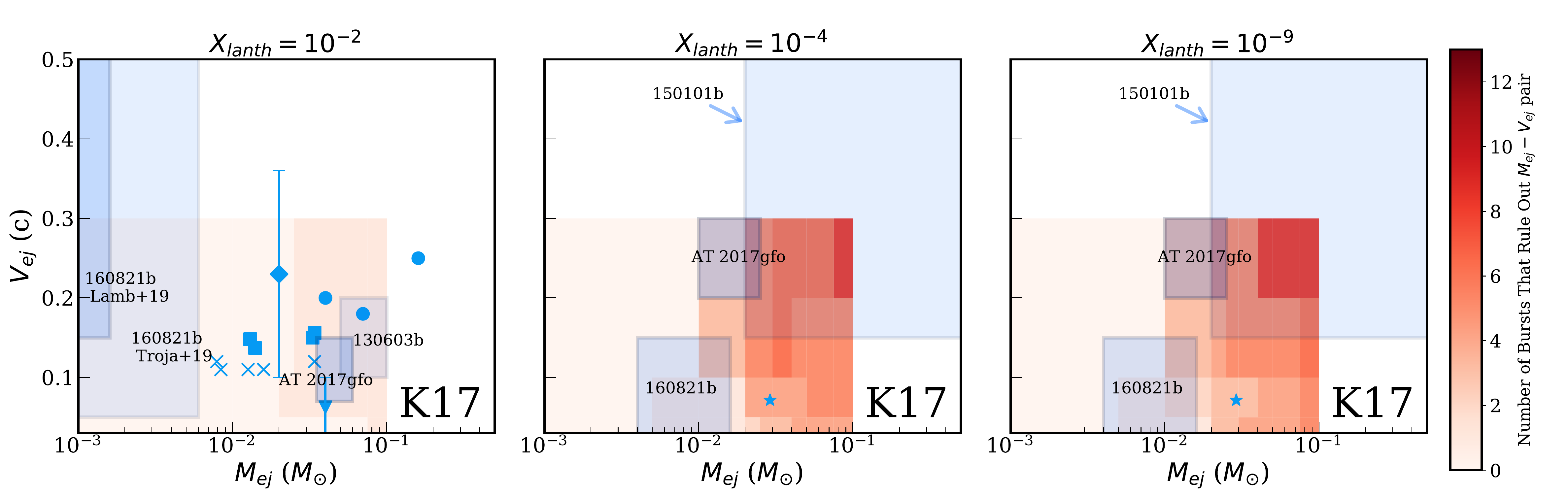}
\vspace{-0.2in}
\caption{Constraints on the $M_{\rm ej}-V_{\rm ej}$ parameter space using upper limits from 14 short GRBs for the M19 (top; \citealt{metzger19}) and the K17 (bottom; \citealt{kasen+17}) model grids. White space on the bottom row grids indicates that the K17 models do not extend to this parameter space. The red colormap corresponds to the number of short GRBs for which an upper limit rules out a kilonova parameterized by the corresponding $M_{\rm ej} - V_{\rm ej}$ pair. From left to right, the progression from redder to bluer components (values of $Y_{\rm e}$ and $X_{\rm lan}$ denoted) shows the larger constraining power of our catalog on a blue component, a product of the nature of our short GRB observations. The estimated ejecta values of 3 short GRB kilonova candidates (130603B; \citealt{barnes+16}, 150101B; \citealt{troja+18}, and 160821B; \citealt{lamb+19,troja+19}), as well as AT~2017gfo, are shown as blue regions that span the error bars of each analysis, and are split into components when relevant. We also plot the dynamical ejecta parameters and the post-merger ejecta parameters of the models used in our ratio analysis (\citealt{Kawaguchi+20} and \citealt{kasen+15}), and the values for a suite of BNS and NSBH kilonova models of \citet{Rosswog+14,Rosswog+17} in their corresponding panels.
\label{fig:kn_param}}
\end{figure*}

\subsection{Quantitative Constraints on Ejecta Properties}

Since each of the K17 and M19 light curves are parameterized by a $M_{\rm ej}-V_{\rm ej}$ pair, we are able to use limits to constrain the allowed $M_{\rm ej}-V_{\rm ej}$ space of each burst's kilonova. For each of the fixed values of $Y_{\rm e}$ or $X_{\rm lan}$, we iterate through the model light curves to determine which $M_{\rm ej}-V_{\rm ej}$ pairs are allowed or not allowed given the luminosity constraints imposed by each limit. If a given burst has multiple constraining observations, we determine the parameter space that is ruled out for each limit separately, and then combine them into a single contour. In this manner, a given $M_{\rm ej}-V_{\rm ej}$ pair only has be to ruled out once for a given burst in order to be ruled out entirely. In Figure~\ref{fig:kn_param} we show the resulting constraints of the 14 short GRBs considered on the K17 and M19 $M_{\rm ej}-V_{\rm ej}$ parameter spaces. In Table~\ref{tab:mej}, we list the median upper limit on the ejecta mass of constraining $M_{\rm ej} - V_{\rm ej}$ pairs for each burst and component. As $V_{\rm ej}$ does not vary much over the range of ejecta masses considered, we do not make constraints as it would be almost entirely based on the grid size (Figure~\ref{fig:kn_param}).

In the red ($X_{\rm lan} = 10^{-2}$) K17 grid, we find upper limits of only 2 bursts (GRBs\,061201 and 160624A) constrain the parameter space, limiting a red component to $M_{\rm ej} < 0.06 M_{\odot}$. The K17 grid constraints of GRB\,160624A ($M_{\rm ej} \lesssim 0.04\,M_{\odot}$) are consistent with those of a separate analysis \citep{O'Connor+21}. For blue and `extremely blue' kilonova components ($Y_{\rm e} = 0.40$ or $X_{\rm lan} = 10^{-4}$ and $Y_{\rm e}=0.45$ or $X_{\rm lan} = 10^{-9}$, respectively), the number of limits which enter the parameter space increases to 9 and 12 bursts respectively, a natural consequence of the prevalence of deep, optical follow-up observations.
The M19 grids overall give similar inferred properties (albeit less constraining), and 13 bursts are able to rule out a fast ($V_{\rm ej} \gtrsim 0.4c$), massive ($M_{\rm ej} \gtrsim 0.4 M_{\odot}$) `extremely blue' ($Y_{\rm e} = 0.45$) component (Table~\ref{tab:mej} and Figure~\ref{fig:kn_param}). Depending on the precise model, meaningful constraints span $M_{\rm ej} \lesssim 0.01-0.1\,M_{\odot}$.

\subsection{Comparisons to Known Candidate Kilonovae and Simulation Parameters}

For context, we include inferred ejecta property values from AT~2017gfo, kilonova candidates and simulations on Figure~\ref{fig:kn_param}. To determine the location of AT~2017gfo on the K17 grids, we use the inferred $M_{\rm ej} - V_{\rm ej}$ parameters compiled in \citet{siegel19} based on optical and NIR observations \citep{kilpatrick+17,villar+17,Cowperthwaite+17,kasen+17,Drout+17,Nicholl+17,Kasliwal+17,Smartt+17,Troja+17,Pian+17, McCully+17,Arcavi+17,Chornock+17,Perego+17,Coughlin+18}. These parameters span $V_{\rm ej} \approx 0.2 - 0.3c$ and $M_{\rm ej} \approx 0.01 - 0.02 M_{\odot}$ for a red component ($X_{\rm lan} \lesssim 10^{-4}$) and $V_{\rm ej} \approx 0.07 - 0.14c$ and $M_{\rm ej} \approx 0.04 - 0.06 M_{\odot}$ for a blue component ($X_{\rm lan} \approx 10^{-2} - 10^{-1}$). To compare AT~2017gfo to the M19 grids, we select the best-fit model using $\chi^2$-minimization between the interpolated master light curves for AT~2017gfo and the grid of models for each $Y_{\rm e}$ value and wavelength (Figure~\ref{fig:kn_param}). The ejecta properties for AT~2017gfo are determined in this manner to ensure a fair comparison to short GRBs in the context of this model grid.

We also place our ejecta property constraints from our short GRB limits in the context of inferred values of candidate short GRB kilonovae. We caution that exact values are subject to variations in data reduction methods and assumed afterglow and kilonova models. In total, the estimates span $0.001 M_{\odot} \lesssim M_{\rm ej} \lesssim 0.1 M_{\odot}$ and $0.025c \lesssim V_{\rm ej} \lesssim 0.9c$ \citep{barnes+16,troja+18,lamb+19,troja+19}. Analyses which describe the predicted ejecta mass as either red or dynamical ejecta are compared to the red component, while those that describe the ejecta as post-merger outflow or accretion disk winds are compared to our bluer components (e.g., \citealt{metzgerfernandez14, perego+14}). Compared to the inferred ejecta properties of AT~2017gfo and GRB\,130603B, only a few bursts have limits which translate to comparable or lower values of $M_{\rm ej}$ (for a given $V_{\rm ej}$).

Finally, we include theoretical predictions for the NSBH and infinite lifetime NS remnant kilonovae models considered earlier \citep{kasen+15,Kawaguchi+20}, as well as additional BNS and NSBH models which span a broad range of input parameters and physics \citep{Rosswog+14, Rosswog+17}. As the NSBH model's post-merger ejecta comprises $Y_{\rm e} = 0.1 - 0.3$ material, we plot both the dynamical and post-merger ejecta parameters on the red component grids only \citep{Kawaguchi+20}. Since the infinite lifetime NS remnant model parameters do not include $Y_{\rm e} < 0.25$ material \citep{kasen+15}, we include this only on the bluer component grids. These three comparison sets, AT~2017gfo, short GRB kilonovae, and inferred values from simulations, are all plotted in Figure~\ref{fig:kn_param}.

This comparison yields several new constraints, although the M19 and K17 grids give slightly different results in terms of direct comparisons (model dependencies further explored in Section~\ref{sec:moddependencies}). Concentrating on the ``red'' component ejecta ($Y_{\rm e} = 0.15$, $X_{\rm lan}=10^{-2}$), the M19 models cannot rule out a red kilonova with properties similar to AT~2017gfo or any short GRB kilonovae, while the single GRB\,160624A rules out such a kilonova in the K17 grid. Furthermore, this same limit on the K17 grid is deep enough to rule out the parameter space occupied by both GRB\,130603B and by the NSBH models given in \citet{Rosswog+17}. For the ``blue'' component ($Y_{\rm e} = 0.40$, $X_{\rm lan}=10^{-4}$, 
upper limits of 5 bursts (GRBs\,050509B, 061201, 080905A, 130822A, 160624A) rule out most of the parameter space allowed by both the upper limit of GRB\,150101B and estimated for AT~2017gfo, compared to $\sim$3 (GRBs\,061201, 130822A, 160624A) on the blue M19 grid. We find that 4 limits rule out the infinite lifetime NS remnant kilonova parameters on the `extremely blue' K17 grid, while 2 are rule out this model on the corresponding M19 grid.

Only one other compilation work has uniformly estimated short GRB kilonova ejecta masses to date \citep{Ascenzi+19}. With the exception of their analysis of GRB\,130603B (which is based on a single detection and thus poorly constrained), \citet{Ascenzi+19} found a range of $M_{\rm ej} \approx 0.05$--$0.1\,M_{\odot}$ using detailed modeling of the few {\it detected} short GRB candidate kilonovae and the K17 models. We derive overall stronger constraints using a newer, larger sample of upper limits and the K17 models, finding $M_{\rm ej} \lesssim 0.05 M_{\odot}$ for 6 bursts (Table~\ref{tab:mej}).

\begin{deluxetable*}{llccccccc}
\savetablenum{1}
\tabletypesize{\footnotesize}
\label{tab:mej}
\centering
\tablecolumns{9}
\tabcolsep0.06in
\tablecaption{Median Constraining Upper Limit of Kilonova Ejecta Mass by Burst and Electron or Lanthanide Fraction}
\tablehead {
&
&
\multicolumn{3}{c}{Median Constraining $M_{\rm ej}$ (M19)}	&
&
\multicolumn{3}{c}{Median Constraining $M_{\rm ej}$ (K17)} \\
\cline{3-5}
\cline{7-9}
\colhead {GRB}		&
\colhead {$z$}		&
\colhead {$Y_{\rm e} = 0.15$}		&
\colhead {$Y_{\rm e} = 0.40$}	&
\colhead {$Y_{\rm e} = 0.45$}	&
&
\colhead {$X_{\rm lan} = 10^{-2}$}		& 
\colhead {$X_{\rm lan} = 10^{-4}$}	&
\colhead {$X_{\rm lan} = 10^{-9}$} \\
&
&
\colhead {($M_{\odot}$)}		&
\colhead {($M_{\odot}$)}	&
\colhead {($M_{\odot}$)}	&
&
\colhead {($M_{\odot}$)}		&
\colhead {($M_{\odot}$)}	&
\colhead {($M_{\odot}$)}
}
\startdata
050509B & 0.225 & - & 0.11 & 0.02 & & - & 0.02 & 0.01 \\
050906 & 0.5$^{\dagger}$ & - & - & 0.24 & & - & - & 0.08 \\
060502B & 0.287 & - & - & 0.33 & & - & 0.06 & 0.04 \\
061201 & 0.111 & - & 0.07 & 0.02 & & 0.06 & 0.02 & 0.01 \\
080503 & 0.5$^{\dagger}$ & - & - & 0.40 & & - & 0.1 & 0.08 \\
080905A & 0.122 & - & 0.09 & 0.02 & & - & 0.01 & 0.01 \\
090305 & 0.5$^{\dagger}$ & - & - & 0.40 & & - & - & - \\
091109B & 0.5$^{\dagger}$ & - & - & 0.33 & & - & - & 0.08 \\
100206 & 0.407 & - & 0.40 & 0.09 & & - & 0.03 & 0.02 \\
130822A & 0.154 & - & 0.04 & 0.01 & & - & 0.02 & 0.02 \\
140516 & 0.5$^{\dagger}$ & - & - & 0.40 & & - & 0.06 & 0.03 \\
150120A & 0.46 & - & - & - & & - & - & 0.10 \\
160624A & 0.483 & 0.11 & 0.05 & 0.04 & & 0.03 & 0.02 & 0.04 \\
180805B & 0.661 & - & - & 0.40 & & - & - & - \\
\enddata
\tablecomments{Numbers show the median $M_{\rm ej}$ of constraining $M_{\rm ej} - V_{\rm ej}$ pairs at the fixed $Y_{\rm e}$ or $X_{\rm lan}$ value. The three left columns show constraints on the grid of models created using analytical perscriptions of \citet{metzger19} (M19) and opacities from \citet{Tanaka+19}. The right columns shows constraints on the grid of models described in \citet{kasen+17} (K17). Dashes show bursts with no constraining $M_{\rm ej} - V_{\rm ej}$ pairs at the fixed $Y_{\rm e}$ or $X_{\rm lan}$ value. \\
$^{\dagger}$ Redshift unknown, $z=0.5$ taken as fiducial value in analysis }
\end{deluxetable*}

\subsection{Model-Dependent Constraints on Kilonova Ejecta Masses}
\label{sec:moddependencies}

Our method of translating limits to allowed ejecta masses and velocities demonstrates a high dependence on the assumed model. Indeed, for a given upper limit, the ejecta mass constraints can vary on the order of $0.1 M_{\odot}$ between the the K17 and M19 analyses (see Table~\ref{tab:mej}). While we analyze a larger number of bursts on the M19 grid, this is entirely because of the larger parameter space enabled by a purely analytic versus numerical calculation of kilonova light curves. Overall, while the analytic M19 light curves enable us to have significantly finer sampling and more dynamic range in $M_{\rm ej}-V_{\rm ej}$, we obtain {\it less} stringent constraints for each given burst than with the K17 models. We attribute this discrepancy to differences in the opacity between K17 and M19.  

In the former, opacity is calculated numerically from the ionization state and abundance distribution of $r$-process species whose bound-bound transitions dominate the kilonova spectrum \citep[in the \texttt{Sedona} code, as in][]{Kasen06,Kasen+13,kasen+17}.  Alternatively, M19 assumes the escaping spectrum resembles a blackbody with a constant effective opacity.  Thus the luminosity in each band around and after peak light where our observations are most constraining depends on the effective temperature of the ejecta and thus the equipartition of luminosity into each band.  As nearly all of our observations are far into the Wien tail of the rest-frame kilonova spectrum, even small discrepancies in temperature can lead to large differences in peak magnitude.  Moreover, while K17 performs a full radiative transfer calculation through each layer of the ejecta, incorporating deposition of radioactive energy and opacity \citep[notably, the ``blue'' kilonova was reproduced assuming a steeply-varying radial density and lanthanide abundance profile;][]{kilpatrick+17,kasen+17}, the effects from emission produced in the optically-thin layers are not considered in M19.  Table~\ref{tab:mej} shows the differences between constraints of the same limits on the M19 grid, which employ opacities that vary with the changing composition and ejecta expansion, and the K17 grid which uses opacities based on a full numerical calculation. The discrepancy in allowed ejecta masses and velocities at similar $Y_{\rm e}$ values reinforces the model-dependent nature of our results. In addition to model dependencies, we note that 5 bursts in our sample (Table~\ref{tab:mej}) have unknown redshifts. However, the majority of these bursts have weak ejecta mass constraints ($M_{\rm ej} \lesssim 0.1 M_{\odot}$) and do not affect our strongest conclusions.

An outstanding theoretical challenge is determining the mechanism that produces the fast ($v \sim 0.2-0.3c$), blue ($Y_{\rm e} \gtrsim 0.25$) ejecta needed to replicate the early ($\delta t \lesssim 1$~day), luminous, optical peak of AT~2017gfo (see summary in \citealt{Metzger17:summary} and \citealt{Banerjee+20}).  It is unclear how these emission mechanisms can match the bright, optical detections of kilonova candidates given the timescales required for neutrinos to raise the electron fraction in the dynamical ejecta \citep{Fernandez+19}. Previous studies have examined potential additional energy sources but have been unable to account for the luminosity peak at timescales greater than a few hours \citep{Metzger+15,Kasliwal+17,Piro+18}. Magnetic fields of the remnant NS or from the accretion disk may play a role in increasing the velocity of high $Y_{\rm e}$ ejecta \citep{Metzger+18,Christie+19,Fernandez+19}. Future observations will enhance our understanding of the early, bright optical peak of AT~2017gfo in connection to short GRB kilonovae.

\section{Discussion}
\label{sec:discussion}

\subsection{Implications for $r$-process Production}

One of the most fundamental questions regarding BNS mergers is the fraction of cosmic $r$-process production for which they are responsible, in part a function of how much material they eject into the interstellar or intergalactic medium. Studies of the kilonovae associated with short GRBs are crucial because they enable an estimate of the average $r$-process yield per event. In turn, this can indicate whether BNS mergers, rather than collapsars \citep{Siege+19b} or other phenomena, created the majority of $r$-process elements in the Universe. Combined with the rates of mergers from either GW or GRB observations, and the timescale over which the mass is assembled, one can estimate the total contribution of mergers to $r$-process enrichment in the Universe (e.g., \citealt{Shen+15,Rosswog+18}).

Initially, the relatively high yields inferred from AT~2017gfo, coupled with the volumetric merger event rates inferred from GW170817 of $R_{BNS} = 1540^{+3200}_{-1220}$ Gpc$^{-3}$ yr$^{-1}$ (90\% confidence; \citealt{Abbott+17a}), suggested that if all kilonovae are similar to this event, the observed $r$-process abundance in the Galaxy would be over-produced \citep{kasen+17}. However, the most recent results from the O3a run of LIGO/Virgo find a lower median event rate of $R_{BNS} = 320^{+490}_{-240}$ Gpc$^{-3}$ yr$^{-1}$ (90\% confidence; \citealt{GWTC2}). This rate is still broadly consistent with estimates obtained from population synthesis \citep{Kim+15} and from short GRBs \citep{fong+15}, implying that average yields close to that of AT~2017gfo may in fact be needed to explain all $r$-process production.

If BNS mergers account for all $r$-process material, this sets a constraint on a combination of the required event rates and ejected masses. In practice, this depends on the abundances of $r$-process elements in the ejecta, rather than the total mass of the material ejected. Indeed, \citet{Rosswog+18} demonstrate that the required event rate is comparatively higher if BNS mergers create all $r$-process elements (e.g., both light and heavy $r$-process elements), while the required event rate is lower if BNS mergers only produce elements beyond atomic mass numbers of $A>130$ (the second r-process peak). The lower atomic mass, first peak $r$-process elements are best probed by the blue kilonova component constraints.

Our large observational catalog provides a far more extensive investigation of the ejecta masses than has previously been possible, and divides the inferred ejecta mass constraints by component. While the upper limits of many short GRBs in our catalogue constrain ejecta masses significantly higher than those inferred for AT~2017gfo (Figure~\ref{fig:kn_param}), there are a handful of events for which we can rule out component masses similar to estimates of this event (see also \citealt{Ascenzi+19}). However, we have also shown that the exact values of these parameters are still dependent on the choice of model (Table~\ref{tab:mej}). For the extremely blue component, the inferred limits reach low ejecta masses of $M_{\rm ej}\lesssim 0.03\,M_{\odot}$ for 6 (4) events in our sample under the K17 (M19) model. If the non-detections of kilonovae in a large fraction of our population are taken to be indicative of a lower $r$-process yields in some systems, then other (unobserved) events must have significantly higher $r$-process yields for BNS mergers to be the sole site of $r$-process nucleosynthesis in the Universe. A final consideration is that a significant fraction of short GRBs (and by extension BNS systems) occur at moderate distances (several $\sim$kpc) from their hosts \citep{FongBerger13,Church+11}. Thus, it is likely that not all $r$-process elements produced in BNS mergers become available for subsequent incorporation into stars.  Nevertheless, this catalog motivates future studies which incorporate these factors, as well as a re-examination of our search strategies for kilonovae following short GRBs.

\subsection{Short GRB Kilonovae in the Context of Current Observational Capabilities}
\label{sec:discussion52}

\begin{figure*}[!t]
\centering
\includegraphics[width=\textwidth]{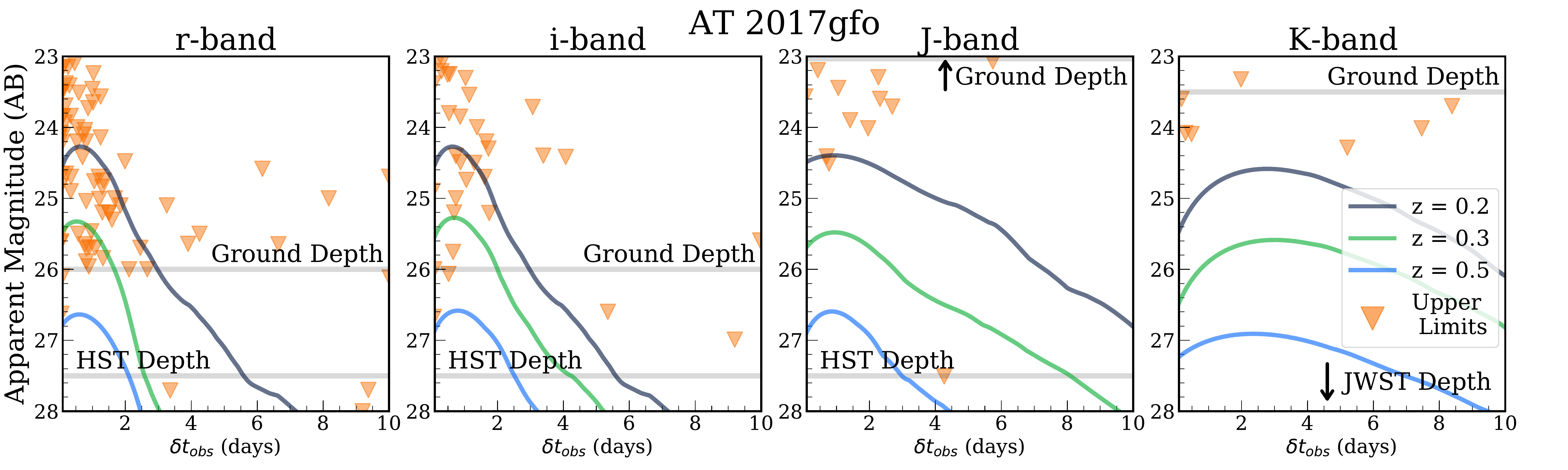}
\includegraphics[width=\textwidth]{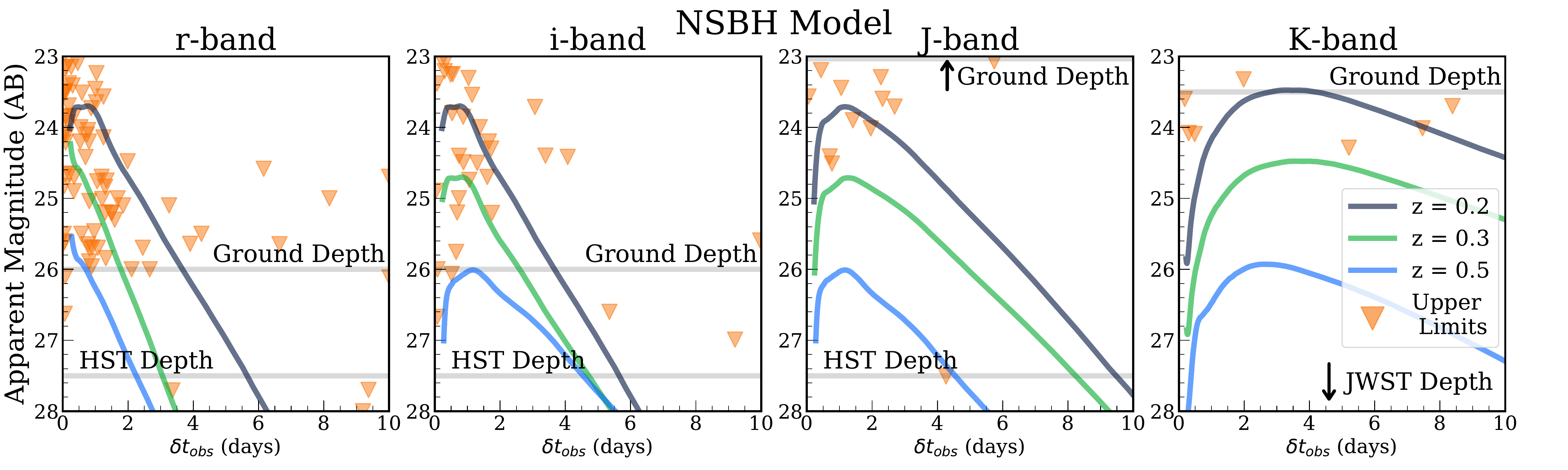}
\vspace{-0.1in}
\caption{Observability of an AT~2017gfo-like  (top) and NSBH (bottom; \citealt{Kawaguchi+20}) kilonova shifted to $z=0.2$, $z=0.3$ and $z=0.5$, the approximate median of the short GRB redshift distribution, in two optical ($ri$-) and two near-infrared ($JK$-) bands. In calculating light curves we account for redshift effects on time and observed band. We overlay historical upper limits as orange triangles and the depths of typical ground- and space-based telescopes as gray bars. In the optical, where ground-based telescopes are more powerful, future nearby ($z \lesssim 0.3$) kilonovae comparable to AT~2017gfo or the NSBH model will be detectable from the ground at $\delta t \lesssim 4$~days. In the near-infrared, space-based follow-up is necessary to detect kilonovae even at optimistic short GRB distances.
\label{fig:kn_obs}}
\end{figure*}

Among our comprehensive sample of 85 short GRBs, we find that only a small fraction of upper limits can place meaningful constraints on kilonova models or the viable $M_{\rm ej}-V_{\rm ej}$ parameter space. To represent where our limits fall with respect to expectations for realistic kilonova emission and current observational capabilities, we compare our catalog of limits to the light curves of AT~2017gfo and the NSBH model of \citet{Kawaguchi+20}. In Figure~\ref{fig:kn_obs} we plot the $riJK$-band light curves of AT~2017gfo and the NSBH model scaled to $z=0.2$, $z=0.3$, and $z=0.5$ (roughly the median detected short GRB redshift), accounting for redshift effects in choosing the relevant model.

Figure~\ref{fig:kn_obs} shows the sheer number of optical ($r$- and $i$-band) upper limits at $\delta t \lesssim 2$~days, which dwarfs the number of deep, NIR observations on similar timescales. However, only 7.9\% of optical ($griz$-bands) upper limits are deep enough and on the relevant timescales to detect a nearby ($z \lesssim 0.3$) AT~2017gfo-like kilonova (Figure \ref{fig:kn_obs}), with the NSBH model faring similarly. Kilonovae have the potential to exhibit redder colors and peak on longer timescales than afterglows, motivating past searches in the NIR bands on $\delta t \gtrsim 0.5$~day. However, only 1.7\% of NIR ($JHK$-bands) upper limits are sufficient to detect the kilonovae of low-redshift ($z \lesssim 0.3$) bursts. More broadly, none of the $\sim 100$ {\it ground-based} NIR observations in our sample uncovered kilonova emission, and in only one case does a ground-based NIR observation probe lower luminosities than  AT~2017gfo (GRB\,160821B; $z=0.1616$; \citealt{troja+19}). Instead, all remaining NIR observations that detected or placed meaningful constraints on kilonova emission involved {\it HST}, which has uncovered kilonova candidates to $z \lesssim 0.6$ \citep{Fox+05,Tanvir+13,berger+13,lamb+19,troja+19,Fong+21}.

Ground-based optical observations have played a comparatively larger role in detecting kilonovae in bluer bands (e.g., GRB\,050709; \citealt{Hjorth+05,Covino+06}, GRB\,060614; \citealt{DellaValle+06}, GRB\,150101B; \citealt{Fong+16}, and GRB\,160821B; \citealt{lamb+19,troja+19}), although the timing of constraining observations must be finely-tuned given the early and bright optical afterglow. Comparing the AT~2017gfo and NSBH kilonova model light curves to the depths of targeted, deep, ground- and space-based observations, it is clear that space-based capabilities are required to detect the ``average'' short GRB at $z=0.5$.

\subsection{Future Short GRB Kilonova Observations}
\label{sec:future_kn}

\begin{figure*}[!t]
\centering
\includegraphics[width=.84\textwidth]{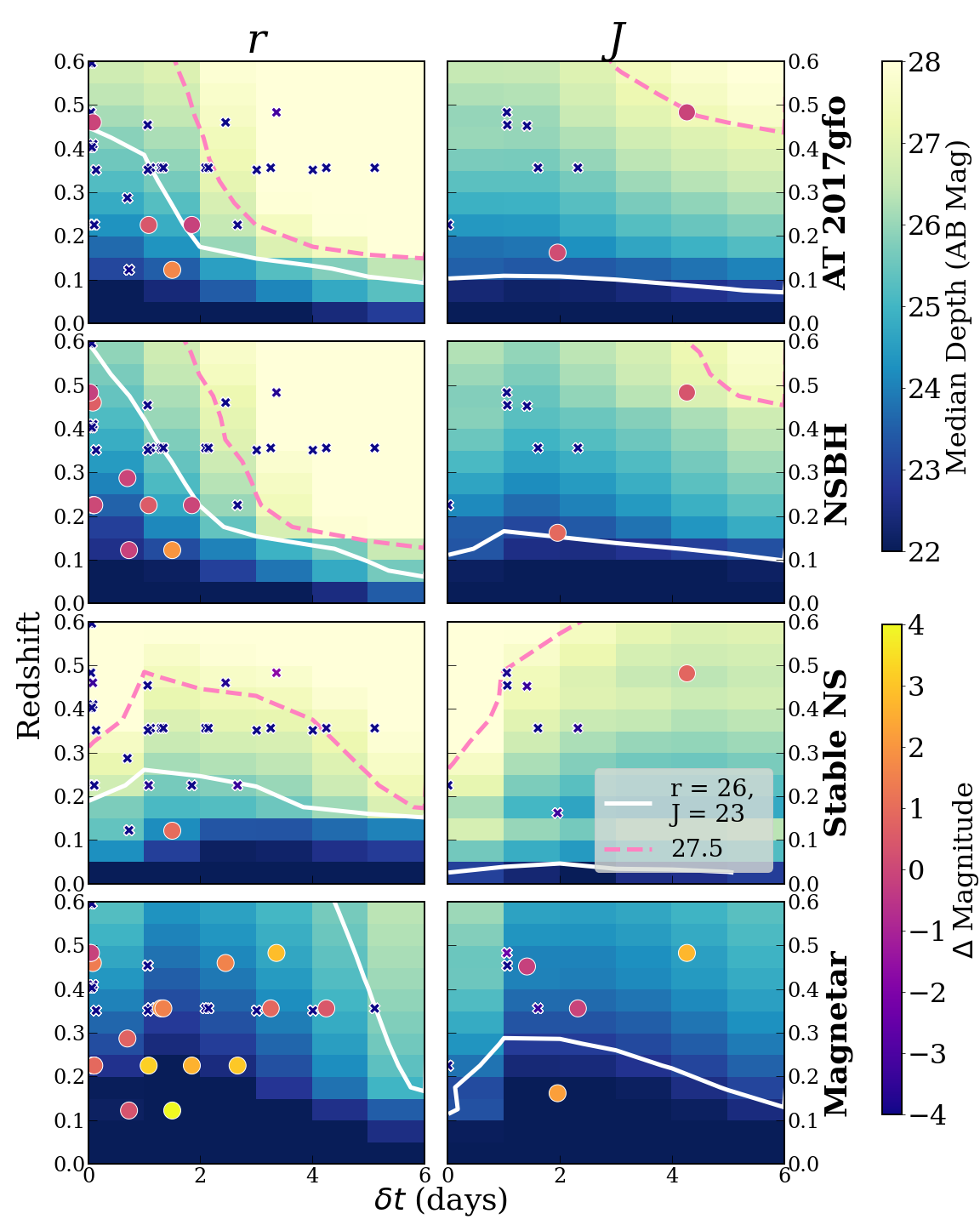}
\caption{The median expected apparent magnitudes of AT~2017gfo light curves and three kilonova models (rows) in the observed $r$- (left column) and $J$-bands (right column) found within 1~day bins at 12 equally spaced redshifts in the range $0 < z < 0.6$. The median depths correspond to the color bar in the upper right corner and account for redshift effects on time and observed wavelength. We also plot two contours showing the limits of ground-based ($m=26$~mag in $r$-band and $m = 23$~mag in $J$-band; while solid lines) and space-based telescopes ($m=27.5$~mag; pink dashed lines). The space below these contours is considered observationally achievable by the respective facilities. We also plot past upper limits of short GRBs with known redshift, dividing them into limits which reached the appropriate depth to rule out the model (circles) and those which were too shallow and did not reach the appropriate depth (crosses). The color of each limit marker corresponds to the difference in magnitude between the observation and the median depth of the model ($\Delta~{\rm Magnitude} = m_{\rm obs} - m_{\rm model}$) and correspond to the color bar in the lower right corner. The majority of observations of low redshift bursts are taken in the $r$-band, despite the NIR light curves' longer timescale of observability.
\label{fig:obsplot_opt}}
\end{figure*}

Motivated by the poor constraining power of most observations in our catalog and the established diversity of kilonovae, we next consider if and how current observing strategies could be adjusted to uncover or place constraints on kilonova emission following short GRBs. In doing this, we also consider the limits of observational resources. Given the precise localizations of short GRBs (as compared to GW events), targeted searches on timescales ranging from $\sim$hours to several days in a variety of filters are possible. However, given the relatively brighter, on-axis afterglows of short GRBs, optical kilonova searches are effectively limited to $\delta t \gtrsim 1$~day. For detections at early times, a determination of the true kilonova contribution would require non-trivial modeling of the bright afterglow, necessitating well-sampled, multi-wavelength observations (for instance, GRB\,160821B; \citealt{lamb+19}).

In this vein, we investigate how the median expected magnitude of AT~2017gfo and kilonova models evolve with $\delta t$ and redshift. We consider the NSBH and infinite lifetime NS remnant previously mentioned in our comparative analysis \citep{villar+17,Kawaguchi+20,kasen+15}. In addition, we include a magnetar-boosted model invoked by \citet{Fong+21} to explain the late-time ($\delta t_{\rm rest} \approx 2.3$~days) infrared detections of GRB\,200522A that were $\approx$10 times the luminosity of AT~2017gfo (but see also \citealt{O'Connor+21} for an alternative interpretation). A magnetar kilonova model has also been invoked to explain the anomalously bright detections of GRBs\,050724A and 070714B \citep{Gao+17}. Though the optimistic infinite lifetime NS remnant and magnetar-boosted kilonovae are expected for a low percentage of cases \citep{MargalitMetzger19}, their frequency has yet to be determined observationally \citep{Schroeder+20}. Thus, probing the luminosities of optimistic models will uniquely enhance our knowledge of the rates of these scenarios. 

For each model in the $grizJK$-bands, we determine the median expected magnitude within 1-day bins for $\delta t \lesssim 6$~days at 12 fixed redshifts equally spaced in the range $0 \lesssim z \lesssim 0.6$. We account for all redshift effects on time and approximate the light curves to the nearest rest-frame band. In Figure~\ref{fig:obsplot_opt}, we show the temporal evolution of each model as a function of redshift in the optical ($r$-band) and NIR ($J$-band). We show the same analysis applied to the $gizK$-bands in the Appendix (Figure~\ref{fig:obsplot_nir}). We also compare the temporal behavior to the depths of targeted searches for kilonova emission. For ground-based searches (solid lines) we compare to limiting magnitudes $m=26$~mag and $m=23$ mag in the $gri$- and $zJK$-bands, respectively, based on a 30-minute exposure time with an 8m-class telescope. Finally, to highlight how this parameter space is sampled by short GRB observations, we plot all upper limits of short GRBs with known redshift. We divide them into limits which reached the appropriate depth to rule out the model (circles) and those which were too shallow and did not reach the appropriate depth (crosses). The color of the limit markers corresponds to the difference in magnitude between the depth of the observation and the median depth of the model in that bin ($\Delta {\rm Magnitude} =  m_{\rm obs} - m_{\rm model}$, where positive values indicate sufficient depths to rule out the model; Figure~\ref{fig:obsplot_opt}).

In the optical bands, AT~2017gfo-like and NSBH model-like kilonovae are only detectable by ground-based telescopes on short timescales ($\delta t \lesssim 2$~days) out to redshifts of $z \approx 0.4$, while ground-based NIR observations cannot play a significant role. The exception across all bands and models is for very nearby $z \lesssim 0.15$ short GRBs, which are detectable on longer timescales to $\delta t \approx 5$~days. Overlaying our short GRB catalog also highlights the regions of parameter space that are infrequently sampled but observationally achievable. For example, though the very low redshift ($z < 0.2$) portion of the $r$-band grids is somewhat populated, few NIR observations exist for these rare, nearby bursts, despite the ground-based observability of typical kilonovae. As NIR light curves (and NSBH models in particular) benefit from the slower time evolution of the lanthanide-rich, higher opacity ejecta, observers have $>5$~days to obtain multiple, deep ground-based follow-up of the most nearby ($z < 0.2$) bursts, potentially supplying critical color information. 
Deep ($m > 24.5$), ground-based optical observations of $z\approx 0.3-0.4$ bursts at $\delta t \approx 1-2$~days can play a role in uncovering or placing meaningful constraints on further AT~2017gfo-like and NSBH model-like kilonovae.

Naturally, optical and NIR space-based imaging (1-2 orbits of {\it HST} reaching $m=27.5$ mag; dashed pink lines) significantly widens the achievable parameter space to higher redshifts and longer timescales, when afterglow contamination is not an issue. Figure~\ref{fig:obsplot_opt} highlights the necessity of space-based facilities like {\it HST} and the upcoming {\it JWST}.

Finally, when considering optimistic but rarer models, such as the magnetar-boosted kilonova, ground-based optical telescopes can play a singular role in ruling out this model to $z \approx 0.6$ and beyond at $\delta t \lesssim 4$~days. Ground-based NIR observations can also play a crucial role in supplying color information to $z \approx 0.3$. Combined with the intriguing counterpart of GRB\,200522A \citep{Fong+21}, we find motivation for late ($\delta t \gtrsim 2$~days), deep ($m > 25$~mag) ground-based optical follow-up of short GRBs to fully and effectively explore the range of kilonova emission. 

In summary, our investigation of the future observability of short GRB kilonovae suggests the best detection strategies are to (i) continue to conduct deep, ground-based follow-up at $\delta t = 1-2$~days as this allows us to probe diverse kilonova at a wide range of distances while avoiding significant afterglow contamination, (ii) observe low redshift bursts in the NIR-bands to $\delta t \approx 8$~days, a region of the parameter space that is currently unexplored by observations, and (iii) use NIR space-based telescopes such as {\it HST} and {\it JWST} to follow-up short GRBs at $z \lesssim 1$ for up to $\sim 10$~days post-burst (Figure~\ref{fig:kn_obs}). This final point will require continued rapid-response capabilities of space-based missions.

\section{Conclusion}

We present a new catalog of 261 observations of 85 short GRBs, (39 with known redshifts; Table~\ref{tab:sample}), the largest sample considered to date. This catalog includes over 50 unpublished upper limits of 23 short GRBs. Our analysis definitively establishes the wide diversity of kilonovae, confirming and expanding on the findings of previous works \citep{Gompertz+18,Ascenzi+19,Rossi+20}. Our emphasis on deep upper limits allows us to explore a portion of the short GRB observational catalog that has not been the focus of previous short GRB kilonova studies. Beyond this, we constrain the ejecta masses and velocities of 14 short GRB kilonovae and inform future kilonova-targeted observing strategies. While only a small fraction of the existing catalog places constraints probed by the parameter space of known kilonovae, the large data set is broadly informative for a diverse set of current (and future) kilonova models and observations. Our main conclusions are:

\begin{itemize}
    \item We find upper limits of 11.8\% and 15.3\% of short GRBs (including 20.5\% and 23.1\% of bursts with known redshift) in our catalog that probe luminosities deeper than those of AT~2017gfo or a fiducial, pole-on NSBH model, respectively. These limits were derived from observations at $0.1 \lesssim \delta t \lesssim 4$~days, reinforcing the need for continued follow up of localized short GRBs within a few days of discovery.
    \item We consider additional models, including that of an indefinitely stable NS remnant kilonova, GW190425-like kilonova, and magnetar remnant-boosted kilonova. We find that $\approx 50\%$ of the upper limits in the short GRB sample probe luminosities lower than the most optimistic model considered (magnetar-boosted).
    \item In general, the kilonova candidates detected from short GRBs are on the upper end of the luminosity distribution, consistent with expectations for the magnitude-limited sample we have constructed.
    \item While we could not place numerical constraints on $V_{\rm ej}$, we can constrain the blue component ejecta masses ($X_{\rm lan}=10^{-4}$) for 9 bursts to $M_{\rm ej}<0.01-0.1 M_{\odot}$. For an extremely blue component ejecta mass ($X_{\rm lan}=10^{-9}$), the limits also constrain $M_{\rm ej}<0.01-0.1 M_{\odot}$ for 12 bursts. We find that precise values are model-dependent.
    \item Despite the sustained NIR-band luminosity of kilonovae like the NSBH merger model, the timescales of past ground-based NIR observations are typically too early to rule out these kilonova models. Our analysis finds motivation for deep ($m > 23$) observations at $\gtrsim 2$~days of nearby ($z\lesssim0.2$) bursts to detect potential NSBH kilonovae.
    \item Despite the sustained optical brightness of the infinite lifetime NS remnant model, few current observations are on the appropriate timescales to probe this model. Future optical follow-up should concentrate on timescales of $\gtrsim 2-6$~days when such models reach their peak. 
    \item Optical observations generally probe realistic kilonova luminosity functions, but there are almost no low-redshift short GRB NIR observations, despite their detectability with current instruments. In addition, ground-based optical telescopes are capable of probing diverse kilonova outcomes for bursts over a range of redshifts, while future space-based facilities will extend the reach of detecting kilonovae to $z\approx 1$, highlighting the importance of rapid-response observations on current and future missions.
\end{itemize}

The past 15 years of well-localized short GRBs discovered by {\it Swift} and concerted follow-up of their afterglows provide a unique opportunity to re-cast deep limits in terms of known kilonovae and a diverse set of optical and NIR models. Positively identifying these transients as kilonovae is difficult at all distances given their faintness and rapid evolution compared to other transients. While detecting kilonovae following short GRBs pushes the limits of current observatories, searches remain a worthwhile endeavor in the current era of GRB discovery and localization. In particular, it is imperative to maintain broad and flexible search strategies (in terms of timescales, depths and colors) to detect or meaningfully constrain kilonova emission across the diversity of potential outcomes. Although current and future transient surveys may be capable of finding $z>0.1$ kilonovae on their own \citep{Scolnic+17,McBrien20}, the only promising method to positively identify them is by association with short GRBs or GW events. Historically, {\it HST} has played an invaluable role in identifying short GRB kilonova candidates and placing deep limits on emission. Moving forward, this telescope and the forthcoming {\it JWST}, coupled with fine-tuned searches from the ground, represent promising avenues to uncover further diversity in kilonova emission.

\facilities{\textit{Swift}(BAT and XRT), \textit{Fermi} (GBM and LAT), \textit{INTEGRAL} (IBIS and ISGRI), \textit{Suzaku} (WAM), \textit{HST} (ACS, WFC3 and WFPC2), Gemini:Gillett (GMOS), Gemini:South (GMOS), Magellan:Baade (FourStar, IMACS, PANIC), Magellan:Clay (LDSS3), MMT (MMIRS, MMTCam and Binospec)}, UKIRT (WFCAM), WHT (ACAM and LIRIS), BAT (SCORPIO), Blanco (ISPI), CAO:2.2m (CAFOS), Danish 1.54m Telescope (DFOSC), DCT (LMI), GTC (CIRCE and OSIRIS), Keck:I (MOSFIRE, LRIS), LBT (LBC), Max Planck:2.2m (GROND), NOT (ALFSOC), NTT (EFOSC2), OANSPM:HJT (RATIR), Subaru (IRCS and AO188), Swope, TNG (NICS), VLT:Yepun (ISAAC, HAWK-I, FORS1 and FORS2), WIYN (OPTIC)


\software{astropy \citep{Astropy2013,Astropy2018}, Numpy \citep{numpy}, pysynphot \citep{pysynphot}, IRAF \citep{Tody86,tody93}, HOTPANTS \citep{becker15}, Astroalign \citep{beroiz+20}, astrometry.net \citep{lang10}, gwemlightcurves \citep{Coughlin+19,Coughlin+20}.}

\section{Acknowledgements}

\noindent We thank Kyohei Kawaguchi and Claudio Barbieri for supplying their kilonova models. We also thank Dan Kasen and Ben Gompertz for useful discussions. J.~R. acknowledges support from the Northwestern Data Science Fellowship. The Fong Group at Northwestern acknowledges support by the National Science Foundation under grant Nos. AST-1814782 and AST-1909358.

Observations reported here were obtained at the MMT Observatory, a joint facility of the University of Arizona and the Smithsonian Institution (program IDs: 2014C-SAO-9, UAO-S221, 2019C-UAO-G199, 2017A-UAO-S190.) MMT Observatory access was supported by Northwestern University and the Center for Interdisciplinary Exploration and Research in Astrophysics (CIERA).
Based on observations obtained at the international Gemini Observatory (Program IDs GS-2013A-Q-31, GN-2014B-Q-10, GN-2014B-Q-49,GN-2016A-DD-1, GN-2016A-DD-1-85, GS-2016A-DD-1,GS-2018A-Q-127, GN-2018A-Q-121), a program of NOIRLab, which is managed by the Association of Universities for Research in Astronomy (AURA) under a cooperative agreement with the National Science Foundation on behalf of the Gemini Observatory partnership: the National Science Foundation (United States), National Research Council (Canada), Agencia Nacional de Investigaci\'{o}n y Desarrollo (Chile), Ministerio de Ciencia, Tecnolog\'{i}a e Innovaci\'{o}n (Argentina), Minist\'{e}rio da Ci\^{e}ncia, Tecnologia, Inova\c{c}\~{o}es e Comunica\c{c}\~{o}es (Brazil), and Korea Astronomy and Space Science Institute (Republic of Korea).
The United Kingdom Infrared Telescope (UKIRT) was supported by NASA and operated under an agreement among the University of Hawaii, the University of Arizona, and Lockheed Martin Advanced Technology Center; operations are enabled through the cooperation of the East Asian Observatory. We thank the Cambridge Astronomical Survey Unit (CASU) for processing the WFCAM data and the WFCAM Science Archive (WSA) for making the data available. Program IDs include U/15A/UA02A, U/16A/UA02A, U/16B/UA03A, U/17A/UA08, and U/18A/UA01. 
Based on observations made with the NASA/ESA Hubble Space Telescope, obtained from the data archive at the Space Telescope Science Institute. STScI is operated by the Association of Universities for Research in Astronomy, Inc. under NASA contract NAS 5-26555.

This publication makes use of data products from the Two Micron All Sky Survey, which is a joint project of the University of Massachusetts, Amherst and the Infrared Processing and Analysis Center/California Institute of Technology, funded by the National Aeronautics and Space Administration and the National Science Foundation.

\appendix 
\restartappendixnumbering

\section{Supplementary Information} \label{appendix}

\startlongtable
\tabletypesize{\footnotesize}
\begin{deluxetable*}{lclcCCCRc}
\centering
\tablecolumns{9}
\tabcolsep0.07in
\tablewidth{0pc}
\tablecaption{Catalog of Short GRB Kilonovae Observations}
\label{tab:sample}
\tablehead {
\colhead {GRB}		&
\colhead {$z$}		&
\colhead {Telescope/Instrument}	&
\colhead {$\delta t$}	&
\colhead {Filter}	&
\colhead {Magnitude}	&
\colhead {Error}	&
\colhead {$\nu L_{\nu}$}	&
\colhead {Reference}	
\\
\colhead {}		&
\colhead {}		&
\colhead {}	&
\colhead {(days)}	&
\colhead {}        &
\colhead {(AB mag)}	&
\colhead {(AB mag)}	&
\colhead {(erg/s)}	&
\colhead {}		
}
\startdata
050509B & 0.225 & WIYN/OPTIC & 0.09 & r & $>$24.2 &   & < 5.5 \times 10^{41} & 2 \\
 &  & WIYN/OPTIC & 0.10 & r & $>$23.8 &   & < 7.8 \times 10^{41} & 2 \\
 &  & WIYN/OPTIC & 0.10 & r & $>$23.9 &   & < 7.7 \times 10^{41} & 2 \\
 &  & WIYN/OPTIC & 0.11 & r & $>$24.1 &   & < 6.1 \times 10^{41} & 2 \\
 &  & Keck-I/LRIS & 1.08 & R & $>$25.6 &   & < 1.5 \times 10^{41} & 3 \\
 &  & VLT/FORS & 1.85 & R & $>$25.1 &   & < 2.4 \times 10^{41} & 4 \\
 &  & 6.0BTA/SCORPIO & 2.67 & R & $>$26.0 &   & < 1.1 \times 10^{41} & 5 \\
050709$^{*}$ & 0.161 & Danish tel./DFOSC & 1.41 & R & 23.0 & 0.1 & 8.1 \times 10^{41} & 6 \\
 &  & VLT/FORS2 & 2.47 & R & 24.0 & 0.1 & 3.2 \times 10^{41} & 7 \\
 &  & Danish tel./DFOSC & 2.50 & R & 23.7 & 0.2 & 4.1 \times 10^{41} & 6 \\
 &  & VLT/FORS1 & 4.36 & V & $>$25.0 & & < 1.6 \times 10^{41} & 7 \\
 &  & $HST$/ACS & 5.60 & F814W & 25.1 & 0.1 & 1.2 \times 10^{41} & 8 \\
 &  & $HST$/ACS & 9.80 & F814W & 25.8 & 0.1 & 6.0 \times 10^{41} & 8 \\
050724A$^{*}$ & 0.257 & Magellan/Baade/PANIC & 1.45 & K & $>$22.1 & & < 1.5 \times 10^{42}  &  9 \\
 &  & VLT/FORS1 & 1.45 & I & 22.8 & 0.1 & 2.2 \times 10^{42} & 10 \\
 &  & Swope 40 in. & 1.53 & I & $>$21.5 &  & < 7.0 \times 10^{41} & 9 \\
 &  & VLT/FORS1 & 1.46 & R & 22.6 & 0.1 & 3.3 \times 10^{42} & 10 \\
 &  & VLT/FORS1 & 3.46 & I & 25.1 & 0.3 & 2.5 \times 10^{41} &  10 \\
050813 & 0.72 & CAHA/CAFOS & 0.55 & I & $>$23.2 &  & < 1.6 \times 10^{43} & 11 \\
 & & CAHA/CAFOS & 0.59 & R & $>$23.5 &  & < 1.6 \times 10^{43} & 11 \\
050906$^{a}$ & 0.5$^{\dagger}$ & VLT/FORS2 & 0.89 & R & $>$26.0 &  & < 7.1 \times 10^{41} & 12 \\
051210 & 0.5$^{\dagger}$ & Magellan/Clay/LDSS3 & 0.81 & r & $>$25.9 &  & 9.3 \times 10^{42} &  13  \\
051221A & 0.546 & Gemini-N/GMOS & 6.16 & r & $>$24.6 &  & < 3.0 \times 10^{42} & 14 \\
060121 & 0.5$^{\dagger}$ & WIYN & 1.40 & R & 24.9 & 0.2 & 1.9 \times 10^{42} & 15 \\
060313 & 0.5$^{\dagger}$ & Gemini-S/GMOS & 10.0 & r & $>$24.7 &  & < 2.2 \times 10^{42} &  13  \\
060502B & 0.287 & Gemini-N/GMOS & 0.70 & R & $>$24.3 &  & < 8.9 \times 10^{41} & 16 \\
060614$^{*}$ & 0.125 & VLT/FORS1 & 3.86 & I & 24.2 & 0.6 & 1.3 \times 10^{41} & 17 \\
 & & VLT/FORS1 & 3.87 & R & 25.5 &  0.3 & 4.6 \times 10^{40} & 17 \\
 & & VLT/FORS1 & 4.85 & R & 25.1 & 0.3  & 6.6 \times 10^{40} & 17 \\
 & & VLT/FORS1 & 6.74 & R & 25.5 &  0.3 &  4.6 \times 10^{40} & 17 \\
 & & VLT/FORS1 & 7.84 & I & 25.1 & 0.4  &  5.7 \times 10^{40} & 17 \\
061201 & 0.111 & VLT/FORS2 & 0.36 & I & 22.8 & 0.1 & 3.6 \times 10^{41} & 18 \\
 & & VLT/FORS2 & 0.38 & R & 23.2 & 0.1 & 3.0 \times 10^{41} & 18 \\
 & & VLT/FORS2 & 1.38 & I & $>$24.0 &  & < 1.2 \times 10^{41}  & 18 \\
 & & VLT/FORS2 & 3.39 & I & $>$24.4 &  & < 8.0 \times 10^{40}  & 18  \\
061217$^{a}$ & 0.827 & Magellan/Clay/LDSS3 & 0.12 & r & $>$23.1 &  & < 3.2 \times 10^{42} &  13  \\
070406$^{a}$ & 0.5$^{\dagger}$ & NOT/ALFSOC & 1.02 & R & $>$23.6 &  & < 5.9 \times 10^{42} & 19 \\
070429B & 0.902 & Gemini-S/GMOS & 0.20 & R & $>$24.7 &  & < 9.8 \times 10^{42} & 20 \\
 & & Blanco/ISPI & 1.13 & J & $>$22.4 &  & < 4.0 \times 10^{43} & 21 \\
070707 & 0.5$^{\dagger}$ & VLT/ISAAC & 2.54 & J & $>$24.5 &  & < 1.4 \times 10^{42} & 22 \\
070714B$^{*}$ & 0.923 & TNG/NICS & 1.00 & J & $>$21.5 &  & < 9.3 \times 10^{43} & 23 \\
& & TNG/NICS & 1.00 & K & 22.8 & 0.3 & 1.7 \times 10^{43} & 23 \\
& & WHT & 1.03 & R & 23.7 & 0.3 & 2.6 \times 10^{43} & 24 \\
& & Keck-I/LRIS & 4.40 & R & 25.7 & 0.3 & 4.1 \times 10^{42} & 25 \\
070724A & 0.457 & Gemini-N/GMOS & 0.10 & g & $>$23.5 &  & < 7.0 \times 10^{42} & 26 \\
070729 & 0.8 & ESO/MPG/GROND & 0.35 & R & $>$24.7 &  & < 7.0  \times 10^{42} & 27 \\
 &  & ESO/MPG/GROND & 0.35 & J & $>$22.7 &  & < 2.3 \times 10^{43} & 27 \\
070809$^{*}$ & 0.473 & Keck-I/LRIS & 0.47 & g & 25.1 & 0.2 & 1.8 \times 10^{42}  & 28 \\
 & & Keck-I/LRIS & 0.47 & R & 23.9 & 0.3 & 4.2 \times 10^{42} & 28 \\
  & & Keck-I/LRIS & 1.46 & g & $>$25.4 &  & < 1.4 \times 10^{42}  & 28 \\
 & & Keck-I/LRIS & 1.46 & R & 24.8 & 0.3 & 1.8 \times 10^{42} & 28 \\
071112B$^{a}$ & 0.5$^{\dagger}$ & Magellan/Clay/LDSS3 & 0.26 & r & $>$23.1 &  & < 9.3 \times 10^{42} & 29 \\
 & & ESO/MPG/GROND & 0.40 & J & $>$21.6 &  & < 2.0 \times 10^{42} & 27 \\
080503 & 0.5$^{\dagger}$ & Gemini-N/GMOS & 0.04 & r & $>$25.6 & & < 9.7 \times 10^{41} & 30 \\
 & & Gemini-N/GMOS & 0.05 & g & 26.5 & 0.2 & 5.5 \times 10^{41} & 30 \\
 & & Keck-I/LRIS & 0.05 & R & $>$25.6 &   & < 9.9 \times 10^{41} & 30 \\
 & & Gemini-N/GMOS& 0.06 & r & $>$26.6 &   & < 3.8 \times 10^{41} & 30 \\
 & & Gemini-N/GMOS & 0.08 & i & $>$26.7 &   & < 3.0 \times 10^{41} & 30 \\ 
 & & Gemini-N/GMOS & 0.09 & z & $>$25.9 &   &  < 5.2 \times 10^{41} & 30 \\
 & & Gemini-N/GMOS & 1.08 & r & 25.3 & 0.1 & 1.3 \times 10^{42} & 30 \\
 & & Gemini-N/GMOS & 1.98 & r & 25.5 & 0.2 &  1.1 \times 10^{42} & 30 \\
 & & Gemini-N/GMOS & 2.09 & g & 26.3 & 0.3 & 7.1 \times 10^{41} & 30  \\
  & & Gemini-N/GMOS & 3.08 & r & 25.7 & 0.3 &  1.1 \times 10^{42} & 30 \\
 & & Gemini-N/GMOS & 4.05 & r & 26.1 & 0.2 & 7.1 \times 10^{41} & 30 \\
 & & $HST$/WFPC2 & 5.36 & F606W & 26.9 &  0.2 &  3.0 \times 10^{41} & 30 \\
080905A & 0.122 & NOT/ALFSOC & 0.35 & R & 24.3 & 0.5 & 1.4 \times 10^{41} & 31 \\
 & & VLT/FORS2 & 0.60 & R & 24.5 & 0.3 & 1.1 \times 10^{41} & 31 \\
 & & ESO/MPG/GROND & 0.73 & r & $>$22.5 &   & < 8.7 \times 10^{41} & 27 \\
 & & VLT/FORS & 1.50 & R & $>$25.2 &   & < 5.7 \times 10^{40} & 31 \\ 
090305$^{a}$ & 0.5$^{\dagger}$ & Gemini-S/GMOS & 0.90 & r & $>$25.7 &   & < 3.7 \times 10^{40} & 32 \\
090515 & 0.403 & Gemini-N/GMOS  & 0.07 & r & 26.3 & 0.1 & 3.1 \times 10^{41} & 33 \\
 & & Gemini-N/GMOS & 1.04 & r & 26.5 & 0.3 & 2.6 \times 10^{41} & 33 \\ 
090621B & 0.5$^{\dagger}$ & RTT150/TFOSC & 0.03 & r & $>$23.2 &  & < 8.9 \times 10^{42} & 34 \\
091109B & 0.5$^{\dagger}$ & VLT/FORS2 & 0.25 & R & 24.3 & 0.1 & 3.2 \times 10^{42} & 32 \\
 &  & VLT/HAWK-I & 0.30 & K & $>$24.1 &  & < 1.1 \times 10^{42} & 32 \\
 &  & VLT/HAWK-I & 0.35 & J & $>$22.9 &  & < 5.7 \times 10^{42} & 32 \\
 & & VLT/FORS2 & 0.43 & R & 24.6 & 0.2 & 2.4 \times 10^{42} & 32  \\
 & & VLT/FORS2 & 1.32 & R & $>$25.8 &   & < 8.0 \times 10^{41} & 32  \\ 
100206 & 0.407 & ESO/MPG/GROND & 0.49 & i & $>$23.3 &  & < 4.0 \times 10^{42} & 27 \\
 &  & Gemini-N/GMOS & 0.65 & i & $>$25.8 &   & < 4.3 \times 10^{41} & 35 \\
100625A & 0.452 & ESO/MPG/GROND & 0.51 & g & $>$23.6 &  & < 6.0 \times 10^{42} & 27 \\
 &  & Magellan/PANIC & 1.41 & J & $>$23.9 &   & < 1.9 \times 10^{42} & 27 \\
100628A & 0.5$^{\dagger}$ & Gemini-N/GMOS & 0.05 & i & $>$24.1 &  & < 3.9 \times 10^{42} & 36 \\
 &  & Magellan/PANIC & 0.74 & J & $>$22.1 &  & < 1.3 \times 10^{43} & 36 \\
100702A & 0.5$^{\dagger}$ & ESO/MPG/GROND & 0.07 & r' & $>$23.2 &  & < 9.1 \times 10^{42} & 27 \\
101219A & 0.718 & Gemini-S/GMOS & 0.04 & i & $>$24.9 &   & < 3.5 \times 10^{42}& 37 \\
 &  & Magellan/FourStar & 0.05 & J & $>$23.6 &   & < 7.9 \times 10^{42} & 37 \\
110420B$^{a}$ & 0.5$^{\dagger}$ & Magellan/IMACS & 0.44 & r & $>$23.5 &   & < 6.5 \times 10^{42} & 36 \\
111020A & 0.5$^{\dagger}$ & Gemini-S/GMOS & 0.74 & i & $>$24.4 &  & < 2.4 \times 10^{42} & 38 \\
 & & VLT/HAWK-I & 0.77 & J & $>$24.5 & & < 2.4 \times 10^{42}  & 32 \\
111117A & 2.211 & GTC/OSIRIS & 0.34 & r & $>$24.9 &   & < 7.1 \times 10^{43} & 39 \\
 &  & Gemini-S/GMOS & 0.57 & r & $>$25.5 &   & < 4.1 \times 10^{43} & 40 \\
120521A & 0.5$^{\dagger}$ & ESO/MPG/GROND & 0.78 & r & $>$24.0 &   & < 4.1 \times 10^{42} & 41 \\
120817B & 0.5$^{\dagger}$ & LCO/duPont/WFCCD & 1.00 & R & $>$23.5 &   & < 7.0 \times 10^{42} & 42 \\
130313A &  0.5$^{\dagger}$ & Gemini-S/GMOS & 0.62 & i & $>$25.7 & & < 7.3 \times 10^{41} & 1 \\
  &  & NOT/ALFSOC & 0.42 & i & $>$23.4 & & < 6.1 \times 10^{42} & 43 \\
 &  & TNG & 0.56 & r & $>$25.0 & & < 1.7 \times 10^{42} & 44 \\
130515A & 0.5$^{\dagger}$ & Gemini-S/GMOS & 0.03 & r & $>$23.5 &   & < 6.5 \times 10^{42} & 45 \\
130603B$^{*}$ & 0.356 & $HST$/ACS & 9.37 & F606W & $>$27.7 &  & < 7.7 \times 10^{40} & 46, 47 \\
 & & $HST$/WFC3 & 9.49 & F160W & 25.8 & 0.2  &  1.4 \times 10^{41} & 46, 47 \\
130716A & 0.5$^{\dagger}$ & Gemini-N/GMOS & 0.81 & r & $>$25.0 &  & < 9.7 \times 10^{41} & 36 \\
130822A & 0.154 & Gemini-N/GMOS & 0.88 & i & $>$24.5 & & < 1.5 \times 10^{41} & 48 \\
& & UKIRT/WFCAM & 5.69 & H & $>$22.8 & & < 3.4 \times 10^{41} & 1 \\
& & UKIRT/WFCAM & 6.69 & J & $>$22.0 & & < 9.4 \times 10^{41} & 1 \\
130912A &  0.5$^{\dagger}$ & WHT/ACAM & 0.84 & g & $>$24.1 & & < 2.2 \times 10^{42} & 49 \\
&  & WHT/LIRIS & 0.88 & J & $>$22.1 & & < 1.3 \times 10^{43} & 1 \\
 &  & HJT/RATIR & 1.04 & r & $>$23.3 & & < 8.5 \times 10^{42} & 50 \\
 &  & HJT/RATIR & 1.04 & i & $>$23.3 & & < 6.5 \times 10^{42} & 50 \\
 &  & HJT/RATIR & 1.04 & J & $>$22.1 & & < 1.3 \times 10^{43} & 50 \\
 &  & UKIRT/WFCAM & 1.2 & J & $>$22.3 & & < 2.1 \times 10^{43} & 1 \\
 &  & UKIRT/WFCAM & 1.2 & H & $>$22.0 & & < 1.1 \times 10^{43} & 1 \\
&  & WHT/LIRIS & 7.71 & J & $>$22.8 & & < 6.8 \times 10^{42} & 1 \\
& & Magellan/FourStar & 11.92 & J & $>$23.7 & & < 2.9 \times 10^{42} & 1 \\
131004A & 0.717 & HJT/RATIR & 0.30 & r & $>$23.4 &  & < 1.7 \times 10^{43} & 51 \\
 &   & HJT/RATIR & 0.30 & i & $>$23.2 &  & < 1.7 \times 10^{43} & 51 \\
 &   & HJT/RATIR & 0.30 & Z & $>$22.7 &  & < 2.3 \times 10^{43} & 51 \\
 &   & HJT/RATIR & 0.30 & Y & $>$22.2 &  & < 3.2 \times 10^{43} & 51 \\
 &   & HJT/RATIR & 0.30 & J & $>$22.5 &  & < 2.0 \times 10^{43} & 51 \\
 &   & HJT/RATIR & 0.30 & H & $>$22.5 &  & < 1.5 \times 10^{43} & 51 \\
131224A$^{a,b}$ & 0.5$^{\dagger}$  & GTC & 1.11 & z & $>$23.1 &  & < 3.1 \times 10^{42} & 52 \\
140402A$^{a}$ & 0.5$^{\dagger}$ & Magellan/Baade/IMACS & 1.21 & r & $>$25.0 &  & < 1.7 \times 10^{42} & 36 \\
140516 & 0.5$^{\dagger}$ & NOT/ALFSOC & 0.08 & R & $>$24.7 &  & < 2.3 \times 10^{42} & 53 \\
 &  & Subaru/IRCS+AO188 & 0.48 & K' & $>$24.1 &  & < 1.1 \times 10^{42} & 54 \\
 &  & Gemini-N/GMOS & 0.52 & i & $>$26.1 &  & < 5.2 \times 10^{41} & 36 \\
 &  & HJT/RATIR & 0.54 & J & $>$22.0 &  & < 1.4 \times 10^{42} & 55 \\
140606A$^{a}$ & 0.5$^{\dagger}$ & BTA & 0.39 & V & $>$24.2 &  & < 4.5 \times 10^{42} & 52 \\
 &  & BTA & 0.43 & R_c & $>$26.0 &  & < 7.0 \times 10^{41} &  52  \\
140619B$^{a,c}$ & 0.5$^{\dagger}$ & Magellan/Baade/FourStar & 1.48 & J & $>$22.9 &  & < 6.1 \times 10^{42} & 36 \\
140622A & 0.959 & GTC & 0.78 & r & $>$25.4 &  & < 4.6 \times 10^{42} & 52 \\
140903A & 0.351 & MMT/MMTCam & 7.53 & r &  $>$22.6 & & < 6.8 \times 10^{42} & 1 \\
  &  & MMT/MMTCam & 7.50 & i &  $>$22.9 & & < 4.3 \times 10^{42} & 1 \\
140930B & 0.5$^{\dagger}$ & GTC & 3.14 & r & $>$24.5 &  & < 2.8 \times 10^{42} & 52 \\
141212A & 0.596 & Gemini-N/GMOS & 0.69 & i &  $>$25.2 & & < 1.8 \times 10^{42} & 1 \\
  &  & Gemini-N/GMOS & 1.75 & i &  $>$25.2 & & < 1.8 \times 10^{42} & 1 \\
150101B$^{*}$ & 0.134 & Magellan/Baade/IMACS & 1.66 & r & 23.0 & 0.2 & 2.4 \times 10^{42} & 56 \\
 &  & Magellan/Baade/IMACS & 2.63 & r & 23.5 & 0.3 & 5.2 \times 10^{41} & 56 \\
 &  & Gemini-S/GMOS & 10.71 & r & $>$24.2 & & < 1.7 \times 10^{41} & 56 \\
150120A & 0.46 & Gemini-N/GMOS & 0.08 & i & $>$26.0 & & < 4.6 \times 10^{41}  & 1 \\
 &  & Gemini-N/GMOS & 0.08 & r & $>$26.1 & & < 5.1 \times 10^{41}  & 1 \\
  &  & Gemini-N/GMOS & 0.13 & z & $>$25.4 & & < 6.6 \times 10^{41}  & 1 \\
 &  & Gemini-N/GMOS & 2.16 & r & $>$25.7 & & < 7.4 \times 10^{41} & 1 \\
150423A & 1.394 & Magellan/IMACS & 0.10 & i & 23.6 & 0.1 & 6.5 \times 10^{43}  & 1 \\
  &  & WHT/ACAM & 0.67 & g & $>$25.2 & & < 2.3 \times 10^{43} & 57 \\
 &  & HJT/RATIR & 1.06 & r & $>$24.8 & & < 2.7 \times 10^{43} &  58 \\
 &  & HJT/RATIR & 1.06 & i & $>$24.7 & & < 2.2 \times 10^{43} & 58 \\
 &  & HJT/RATIR & 1.06 & z & $>$22.0 & & < 2.2 \times 10^{43} & 58 \\
150424A & 0.5$^{\dagger}$ & ESO/MPG/GROND & 1.81 & r' & 23.1 & 0.1 & 5.5 \times 10^{43} & 59 \\
 & & ESO/MPG/GROND & 1.81 & i' & 23.0 & 0.1 & 4.7 \times 10^{43} & 59 \\
 & & ESO/MPG/GROND & 1.81 & z' & 22.6 & 0.1 & 5.7 \times 10^{43} & 59 \\
  & & $HST$/WFC3 & 6.64 & F606W & 25.9 & 0.1 & 4.1 \times 10^{42} & 60 \\
 & & $HST$/WFC3 & 6.68 & F125W & 25.3 & 0.1 & 3.7 \times 10^{42} & 60 \\
 & & $HST$/WFC3& 6.71 & F160W & 25.1 & 0.1 & 3.3 \times 10^{42} & 60  \\
 & & $HST$/WFC3 & 9.23 & F606W & 26.9 & 0.1 & 1.6 \times 10^{42} & 60  \\
 & & $HST$/WFC3 & 9.26 & F125W & 26.3 & 0.2 & 1.5 \times 10^{42} & 60  \\
 & & $HST$/WFC3 & 9.30 & F160W & 25.8 & 0.1 & 1.8 \times 10^{42} & 60 \\
150831A &  0.5$^{\dagger}$ & ESO/MPG/GROND  & 0.53 & g' & $>$24.5 & & < 3.6 \times 10^{42} & 61 \\
  &  & ESO/MPG/GROND  & 0.53 & r' & $>$24.2 & & < 3.6 \times 10^{42} & 61 \\
 &  & ESO/MPG/GROND  & 0.53 & i' & $>$23.8 & & < 4.2 \times 10^{42}  & 61 \\
 &  & ESO/MPG/GROND  & 0.53 & z' & $>$23.6 & & < 4.2 \times 10^{42} & 61 \\
 &  & ESO/MPG/GROND  & 0.53 & J & $>$21.3 & & < 2.7 \times 10^{43} & 61 \\
151228A$^{a}$ & 0.5$^{\dagger}$ & GTC & 1.14 & i & $>$23.7 &  & < 5.1 \times 10^{42} & 52 \\
160303A &  0.5$^{\dagger}$ & UKIRT/WFCAM & 0.84 & J & $>$22.0 & & < 1.4 \times 10^{43} & 1 \\
 &  & UKIRT/WFCAM & 0.87 & K & $>$22.0 & & < 7.7 \times 10^{42} & 1 \\
 &  & HJT/RATIR & 0.87 & r & $>$23.7 & & < 5.6 \times 10^{42} & 62 \\
 &  & HJT/RATIR & 0.87 & i & $>$23.9 & & < 4.0 \times 10^{42} &  62 \\
 &  & HJT/RATIR & 0.87 & Z & $>$22.4 & & < 1.3 \times 10^{43} &  62 \\
 &  & HJT/RATIR & 0.87 & Y & $>$21.8 & & < 2.1 \times 10^{43} &  62 \\
 &  & HJT/RATIR & 0.87 & J & $>$21.5 & & < 2.2 \times 10^{43} &  62 \\
 &  & HJT/RATIR & 0.87 & H & $>$21.4 & & < 1.8 \times 10^{43} &  62 \\
 &  & ESO/MPG/GROND & 1.73 & i' & $>$24.3 & & < 2.6 \times 10^{42} & 63 \\
 &  & ESO/MPG/GROND & 1.73 & z' & $>$24.4 & & < 2.0 \times 10^{42} & 63 \\
 &  & MMT/MMTCam & 3.90 & r & $>$25.6 & & < 9.5 \times 10^{41} & 1 \\
  &  & MMT/MMTCam & 10.82 & i & $>$24.5 & & < 2.2 \times 10^{42} & 1 \\
160408A &  0.5$^{\dagger}$ & Gemini-N/GMOS & 0.05 & r & 24.8 & 0.1 & 2.1 \times 10^{42} & 1 \\
  &  & Gemini-N/GMOS & 0.97 & r &  $>$25.5 & & < 1.1 \times 10^{42} & 1 \\
160410A & 1.717 & UKIRT/WFCAM & 1.06 & J & $>$22.2 & & < 1.2 \times 10^{43} & 1 \\
  &  & UKIRT/WFCAM & 1.09 & K & $>$21.8 & & < 9.4 \times 10^{42} & 1 \\
 &  & NOT/ALFSOC & 1.68 & r &  $>$25.0 & & < 1.7 \times 10^{42} & 64 \\
160411A &  0.5$^{\dagger}$ & Gemini-S/GMOS & 0.34 & i & 23.3 & 0.2 & 6.5 \times 10^{42} & 1 \\
 &  & ESO/MPG/GROND  & 0.34 & g' &  $>$23.6 & & < 8.0 \times 10^{42} & 65 \\
 &  & ESO/MPG/GROND  & 0.34 & r' &  $>$23.8 & & < 5.0 \times 10^{42} & 65 \\
 &  & ESO/MPG/GROND  & 0.34 & i' &  $>$23.1 & & < 8.4 \times 10^{42} & 65 \\
 &  & ESO/MPG/GROND  & 0.34 & z' &  $>$22.9 & & < 8.2 \times 10^{42} & 65 \\
 &  & ESO/MPG/GROND  & 0.34 & J &  $>$21.3 & & < 2.7 \times 10^{43} & 65 \\
 &  & ESO/MPG/GROND  & 0.34 & H &  $>$20.8 & & < 3.2 \times 10^{43} & 65 \\
160601A &  0.5$^{\dagger}$ & DCT/LMI & 1.60 & r &  $>$25.0 & & < 1.7 \times 10^{42} & 66 \\
160612A$^{a}$ &  0.5$^{\dagger}$ & UKIRT/WFCAM & 2.69 & J &  $>$23.7 & & < 2.9 \times 10^{42} & 1 \\
  &  & UKIRT/WFCAM & 5.74 & J &  $>$23.1 & & < 5.3 \times 10^{42} & 1 \\
160624A & 0.483 & Gemini-N/GMOS & 0.003 & r & $>$25.5 & & < 1.0 \times 10^{42} & 67 \\
 &  & Gemini-N/GMOS & 0.05 & r & $>$24.8 & & < 1.9 \times 10^{42} & 1 \\
 &  & UKIRT/WFCAM & 1.05 & J & $>$23.4 & & < 3.4 \times 10^{42} & 1 \\
  &  & $HST$/ACS & 3.36 & F606W & $>$27.7 & & < 1.3 \times 10^{41} & 68, 1$^{e}$ \\
 &  & $HST$/WFC3 & 4.27 & F125W & $>$27.5 & & < 8.2 \times 10^{40} & 68, 1$^{e}$ \\
 &  & $HST$/WFC3 & 4.33 & F160W & $>$27.3 & & < 7.4 \times 10^{40} & 68, 1$^{e}$ \\
160821B$^{*}$ & 0.1616 & WHT/ACAM & 1.06 & r & 23.8 & 0.1 &  3.8 \times 10^{41} & 69 \\
 & & WHT/ACAM & 1.08 & z & 23.6 & 0.2 &  3.1 \times 10^{41} & 69 \\
 & & GTC/CIRCE & 1.94 & H & $>$23.8 &   & < 1.5 \times 10^{41} & 70 \\
 & & NOT/ALFOSC & 1.95 & r & 24.8 & 0.1 &  1.5 \times 10^{41} & 69 \\
 & & GTC/CIRCE & 1.96 & J & $>$24.0 &   & < 1.7 \times 10^{41} & 70 \\
 & & NOT/ALFOSC & 1.99 & z & 23.9 & 0.2 &  2.4 \times 10^{41} & 69 \\
 & & GTC/OSIRIS & 2.02 & g & 25.6 & 0.2 &  2.5 \times 10^{41} & 69  \\
 & & GTC/OSIRIS & 2.03 & r & 24.8 & 0.1 &  1.6 \times 10^{41} & 69 \\
 & & GTC/OSIRIS  & 2.04 & i & 24.5 & 0.1 &  1.7 \times 10^{41} & 69 \\
 & & GTC/OSIRIS & 2.04 & z & 24.3 & 0.2 &   1.6 \times 10^{40} & 69 \\
 & & $HST$/WFC3/UVIS & 3.64 & F606W & 25.9 & 0.1 &  5.6 \times 10^{40} & 69 \\
 & & $HST$/WFC3/IR & 3.71 & F160W & 24.4 & 0.1 &  8.4 \times 10^{40} & 69 \\
 & & $HST$/WFC3/IR & 3.76 & F110W & 24.7 & 0.2 &  8.7 \times 10^{40} & 69 \\
 & & GTC/OSIRIS & 3.98 & g & 26.0 & 0.2 &  6.7 \times 10^{40} & 69 \\
 & & GTC/OSIRIS & 4.00 & i & 25.7 & 0.4 &  5.5 \times 10^{40} & 69 \\
 & & GTC/OSIRIS & 4.99 & r & 26.1 & 0.3 &  4.6 \times 10^{42} & 69  \\
 & & GTC/OSIRIS & 6.98 & g & 26.9 & 0.2 &  2.9 \times 10^{40} & 69 \\
 & & GTC/OSIRIS & 9.97 & i & $>$25.6 &   & < 6.0 \times 10^{40} & 69 \\
 & & GTC/OSIRIS & 10.00 & g & $>$25.7 &   & < 8.7 \times 10^{40} & 70 \\
 & & $HST$/WFC3/UVIS & 10.01 & F606W & $>$26.1 &   & < 4.6 \times 10^{40} & 69 \\
 & & $HST$/WFC3/UVIS & 10.40 & F606W & 27.7 & 0.1 &  1.1 \times 10^{40} & 69 \\
 & & $HST$/WFC3/IR & 10.46  & F160W & 26.6 & 0.2 &  1.2 \times 10^{40} & 69 \\
 & & $HST$/WFC3/IR & 10.53 & F110W & 26.7 & 0.2 &   1.4 \times 10^{40} & 69 \\
161001A & 0.891 & ESO/MPG/GROND & 0.12 & J &  $>$21.3 & & < 1.1 \times 10^{44} & 71 \\ 
161104A & 0.65 & Gemini-S/GMOS & 0.72 & r & $>$25.4 &  & < 1.2 \times 10^{42} & 72 \\
170112A &  0.5$^{\dagger}$ & UKIRT/WFCAM &  1.17 & J & $>$21.3 & & < 2.8 \times 10^{43} & 1 \\
170127B &  0.5$^{\dagger}$ & UKIRT/WFCAM & 0.55 & J &  $>$21.0 & & < 3.5 \times 10^{43} & 1 \\
  &  & NOT/ALFSOC  & 0.19 & r &  $>$23.4 & & < 7.7 \times 10^{42} & 73 \\
170428A  & 0.454  & MMT/MMTCam  & 3.07 & i & $>$23.7 & & < 3.6 \times 10^{42} & 1 \\
 &  & MMT/MMTCam & 4.08 & i & $>$24.4 & & < 1.9 \times 10^{42} & 1 \\
 &  & UKIRT/WFCAM & 6.24 & J &  $>$22.0 & &< 1.1 \times 10^{42} & 1 \\
  &  & UKIRT/WFCAM & 7.22 & J &  $>$22.7 & & < 5.9 \times 10^{42} & 1 \\
 &  & UKIRT/WFCAM& 8.24 & J &  $>$22.7 & & < 5.9 \times 10^{42} & 1 \\
 &  & UKIRT/WFCAM & 10.23 & J &  $>$22.5 & & < 7.1 \times 10^{42} & 1 \\
170524A &  0.5$^{\dagger}$ & UKIRT/WFCAM & 0.79 & K &  $>$21.3 & & < 1.5 \times 10^{43}  & 1 \\
  &  & UKIRT/WFCAM & 2.76 & K &  $>$21.5 & & < 1.2 \times 10^{43}  & 1 \\
180418A & 1.0$^{\dagger}$ & Gemini-N/GMOS & 4.80 & r & $>$25.2 &  & < 1.4 \times 10^{42} & 74 \\
180715A$^{a}$ & 0.5$^{\dagger}$ & UKIRT/WFCAM  & 0.52 & J & $>$21.0 & & < 3.4 \times 10^{43} & 1 \\
 &  & Gemini-S/GMOS & 0.54 & r & $>$24.0 & & < 4.3 \times 10^{42} & 1 \\
 &  & UKIRT/WFCAM  & 1.51 & J & $>$20.9 & & < 4.1 \times 10^{42} & 1 \\
 &  & UKIRT/WFCAM  & 2.58 & J & $>$21.7 & & < 1.8 \times 10^{43} & 1 \\
 &  & UKIRT/WFCAM  & 7.54 & J & $>$21.4 & & < 2.4 \times 10^{43} & 1 \\
180718A$^{a, d}$ &  0.5$^{\dagger}$ & Gemini-N/GMOS & 1.50 & r &  $>$25.2 & & < 1.4 \times 10^{42}  & 1 \\
180727A &  0.5$^{\dagger}$ & Magellan-Clay/LDSS3 & 1.66 & i &  $>$24.2 & & < 2.9 \times 10^{42}  & 1 \\
180805B &  0.661 & UKIRT/WFCAM & 0.04 & J & $>$22.3 & & < 2.1 \times 10^{43} & 1 \\
 &  & ESO/HAWK-I   & 0.70 & J & $>$24.4 & & < 3.0 \times 10^{42} & 75 \\
 &  & NTT/EFOSC2   & 0.74 & i & $>$25.0 & & < 2.8 \times 10^{42} & 76 \\
 & & Magellan-Baade/IMACS & 0.80 & r & $>$24.2 & & < 7.1 \times 10^{42} & 1 \\
 &  & ESO/FORS2   & 0.82 & R & $>$25.7 & & < 1.8 \times 10^{42} & 75  \\
 &  & UKIRT/WFCAM   & 1.04 & J & $>$22.1 & & < 2.6 \times 10^{43} & 1 \\
181123B & 1.754 & Keck I/MOSFIRE & 0.43 & J & $>$23.2 &  & < 9.9 \times 10^{43} & 77 \\
 & & MMT/MMIRS & 2.27 & J & $>$23.3 &  & < 9.0 \times 10^{43} & 77 \\
181126A &  0.5$^{\dagger}$ & Gemini-N/GMOS & 0.18 & z & $>$25.3 & & < 8.9 \times 10^{41} & 78 \\
  &  & Keck I/MOSFIRE & 0.18 & K_s & $>$23.6 & & < 1.8 \times 10^{42} & 79 \\
190427A$^{a}$ &  0.5$^{\dagger}$ & NOT/ALFSOC & 0.02 & r & $>$23.5 & & < 6.9 \times 10^{42} & 80 \\
191031D &  0.5$^{\dagger}$ & MMT/Binospec  & 0.18 & r & $>$24.1 & & < 4.1 \times 10^{42} & 1 \\
  &  & MMT/Binospec  & 0.18 & z & $>$22.6 & & < 1.1 \times 10^{43} & 1 \\
 &  & HJT/RATIR  & 0.21 & r & $>$22.8 & & < 1.4 \times 10^{42}  & 81 \\
 &  & HJT/RATIR  & 0.21 & i & $>$23.1 & & < 8.4 \times 10^{42} & 81 \\
 &  & MMT/Binospec & 1.19 & r & $>$24.4 & & < 3.1 \times 10^{42} & 1 \\
 &  & Gemini-N/GMOS & 1.30 & r & $>$24.8 & & < 2.0 \times 10^{42} & 82 \\
200522A$^{*}$ & 0.5537 & Gemini/GMOS & 3.12 & r & 26.0 & 0.4 & 8.8 \times 10^{41} & 68 \\
 & & $HST$/WFC3 & 3.52 & F125W & 24.8 & 0.1 &  1.4 \times 10^{42} &  83 \\
 & & $HST$/WFC3 & 3.66 & F160W & 24.9 & 0.1 &  9.8 \times 10^{41} & 83 \\
200623A$^{a, d}$ &  0.5$^{\dagger}$ & LBT/LBC & 2.11 & r & $>$26.0 & & < 6.9 \times 10^{41} & 84 \\
201221D & 1.045 & MMT/MMIRS & 0.40 & J & $>$23.1 & & < 3.1 \times 10^{43} & 1 \\
 & & MMT/MMIRS & 1.43 & J & $>$23.3 & & < 2.6 \times 10^{43} & 1 \\
\enddata
\tablecomments{This sample includes previously published (i) detections of kilonova candidates (marked with $*$ symbols), (ii) deep ($m \gtrsim 21$ AB mag) upper limits, and (iii) low-luminosity ($\nu L_{\nu} \lesssim 5 \times 10^{41}$ erg s$^{-1}$) afterglow detections. \\
\textbf{References}: (1) - This work, 
(2) - \cite{Bloom+06}, (3) - \cite{GCN3409}, (4) - \cite{Hjorth+05}, (5) - \cite{Castro-Tirado+05}, 
(6) - \cite{Hjorth+05b}, (7) - \cite{Covino+06}, (8) - \cite{Fox+05}, 
(9) - \cite{Berger+05}, (10) - \cite{Malesani+07}, 
(11) - \cite{Ferrero+07}, (12) - \cite{Levan+08}, (13) - \cite{Berger+07}, 
(14) - \cite{Soderberg+06},  (15) - \cite{levan+06}, (16) - \cite{Price+06}, 
(17) - \cite{Jin+15}, (18) - \cite{Stratta+07}, (19) - \cite{Malesani+07b}, (20) - \cite{GCN6774}, (21) - \cite{Nysewander+07}, (22) - \cite{Piranomonte+07}, (23) - \cite{Covino+07}, (24) - \cite{Levan+07_GCN}, (25) - \cite{Perley+07_070714b}, (26) - \cite{Berger+09}, (27) - \cite{Guelbenzu+12}, (28) - \cite{Perley+07}, (29) - \cite{BergerChallis07}, (30) - \cite{Perley09}, (31) - \cite{Rowlinson+10b}, (32) - \cite{Tunnicliffe+14}, (33) - \cite{Rowlinson+10}, (34) - \cite{Galeev+09}, (35) - \cite{Perley+12}, (36) - \cite{fong+15}, (37) - \cite{Fong+13}, (38) - \cite{Fong+12}, (39) - \cite{Sakamoto+13}, (40) - \cite{Margutti+12}, (41) - \cite{Rossi+12}, (42) - \cite{Fong+12-GCN}, (43) - \cite{GCN14301}, (44) - \cite{GCN14307}, (45) - \cite{CenkoCucchiara13}, (46) -  \cite{berger+13}, (47) - \cite{Tanvir+13}, (48) - \cite{GCN15121}, (49) - \cite{GCN15224}, (50) - \cite{GCN15226}, (51) - \cite{Littlejohns+13}, (52) - \cite{pandey+19}, (53) - \cite{GCN16290}, (54) - \cite{Minowa+14}, (55) - \cite{Butler+14}, (56) - \cite{Fong+16},  (57) - \cite{GCN17747}, (58) - \cite{GCN17754}, (59) - \cite{Knust+17}, (60) - \cite{Jin+18},  (61) - \cite{GCN18219}, (62) - \cite{GCN19146}, (63) - \cite{GCN19144}, (64) - \cite{GCN19300}, (65) - \cite{GCN19292}, (66) - \cite{GCN19497}, (67) - \cite{GCN19565}, (68) - \cite{O'Connor+21}, (69) - \cite{lamb+19}, (70) - \cite{troja+19}, (71) - \cite{GCN19975}, (72) - \cite{Nugent+20}, (73) - \cite{GCN20549},  (74) - \cite{RoucoEscorial+20},  (75) - \cite{GCN23086}, (76) - \cite{GCN23090}, (77) - \cite{Paterson+20}, (78) - \cite{GCN23458}, (79) - \cite{GCN23462}, (80) - \cite{GCN24265}, (81) - \cite{GCN26121}, (82) - \cite{GCN26147}, (83) - \cite{Fong+21}, (84) - \cite{GCN28034}. \\
\textbf{Redshift References}:
\cite{Bloom+06}, 
\cite{Berger+05}, 
\cite{Prochaska+06}, 
\cite{Soderberg+06}, 
\cite{Bloom+07}, 
\cite{GCN5275}, 
\cite{Stratta+07}, 
\cite{GCN5965}, 
\cite{Cenko+08}, 
\cite{Berger+09}, 
\cite{Nugent+20}, 
\cite{Berger10}, 
\cite{Rowlinson+10b}, 
\cite{Rowlinson+10}, 
\cite{Perley+12}, 
\cite{Fong+13}, 
\cite{Selsing+18}, 
\cite{GCN15178_Wiersema}, 
\cite{GCN15307}, 
\cite{Troja+2016}, 
\cite{GCN17177}, 
\cite{GCN17358}, 
\cite{Selsing+19} 
\cite{fong+15}, \cite{Fong+16}, \cite{GCN19565}, 
\cite{lamb+19}, \cite{troja+19}, \cite{Nugent+20}, \cite{GCN21059}, 
\cite{RoucoEscorial+20}, 
This work (GRB\,180805B), 
\cite{Paterson+20}, \cite{Fong+21}, \cite{GRB201221D_GCNz}. 
\\
Unless noted, all bursts are detected by \textit{Swift}-BAT and \textit{Swift}-XRT. \\
$^{\dagger}$ Redshift unknown, $z=0.5$ taken as fiducial value in analysis, except for GRB\,180418A which is assumed to be $z=1.0$ (c.f., \citealt{RoucoEscorial+20}). \\
$^{*}$ Kilonova candidate \\
$^{a}$ Burst does not have an X-ray afterglow found by \textit{Swift}-XRT. \\
$^{b}$ Burst detected by \textit{INTEGRAL}-IBIS/ISGRI, not detected by \textit{Swift}-BAT. \\
$^{c}$ Burst detected by \textit{Fermi}-GBM, \textit{Fermi}-LAT and by \textit{Suzaku}-WAM, not detected by \textit{Swift}-BAT. \\
$^{d}$ Limit applies to partial coverage of BAT localization \\
$^{e}$ We present a separate analysis of observations published in (68), finding upper limits deeper by 0.2--0.3 magnitudes.}
\end{deluxetable*}

\begin{figure*}
\centering
\includegraphics[width=\textwidth]{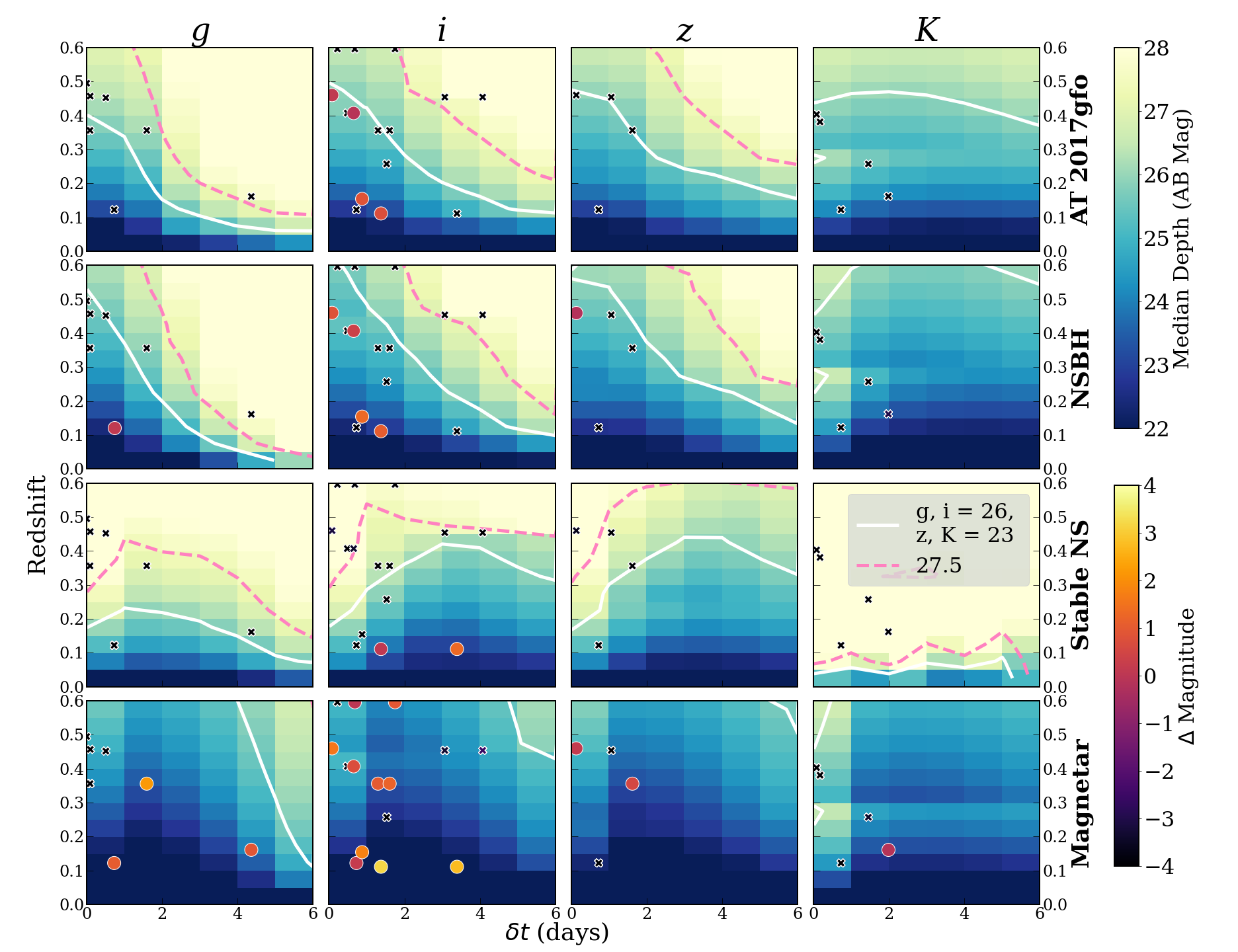}
\caption{The same analysis as Figure~\ref{fig:obsplot_opt} and described in Section~\ref{sec:future_kn} but shown in the observed $g$-, $i$-, $z$- and $K$-bands. As also demonstrated by Figure~\ref{fig:obsplot_opt}, the optical bands are generally more populated with observations of low redshifts bursts than the NIR.
\label{fig:obsplot_nir}}
\end{figure*}

\bibliographystyle{aasjournal}
\bibliography{refs}

\end{document}